\documentclass[useAMS]{mn2e}
\usepackage{amssymb}
\usepackage{graphicx}
\begin{document}

\title[The regeneration of stellar bars by tidal interactions]
  {The regeneration of stellar bars by tidal interactions. Numerical
   simulations of fly-by encounters.}

\author[I.~Berentzen et al.]
       {I.~Berentzen$^{1,2}$\thanks{Email: iberent@uni-sw.gwdg.de},
        E.~Athanassoula$^2$,
        C.H.~Heller$^3$, and
        K.J.~Fricke$^1$ \\
        $^1$Universit\"ats--Sternwarte, Geismarlandstra\ss e 11,
            D-37083 G\"ottingen, Germany\\
        $^2$Observatoire Astronomique de Marseille-Provence, 2 Place
            Le Verrier, F-13248 Marseille Cedex 4, France\\
        $^3$Georgia Southern University, Department of Physics,
            Statesboro, GA 30460, U.S.A.\\
       }

\maketitle

 
\begin{abstract}
 We study the regeneration of stellar bars triggered by a tidal
 interaction, using numerical simulations of either purely stellar or
 stellar+gas disc galaxies. We find that interactions which are
 sufficiently strong to regenerate the bar in the purely stellar
 models do not lead to a regeneration in the dissipative models,
 owing to the induced gas inflow in those models. In models in which
 the bar can be regenerated, we find a tight correlation between the
 strength and the pattern speed of the induced bar. This relation can
 be explained by a significant radial redistribution of angular
 momentum in the disc due to the interaction, similar to the
 processes and correlations found for isolated barred spirals.
 We furthermore show that the regenerated bars show the same
 dynamical properties as their isolated counterparts.
\end{abstract}



\section{Introduction} \label{intro}
 
 Barred galaxies amount to more than one third of the catalogued disc
 galaxies in optical wavelengths (e.g., de Vaucouleurs 1963) and to
 about two thirds in the near-infrared \cite{esk00}. It has been
 established by now, that the {\em life} of a bar can be divided into
 several episodes: its formation, evolution, dissolution and maybe
 its regeneration (e.g., see review by Friedli 1999).

 Two mechanisms for the formation of bars in disc galaxies are being
 widely accepted at present. Bar formation can occur, as shown in
 many numerical simulations, spontaneously by a global instability in
 cold, rotationally supported stellar discs (e.g., Miller,
 Prendergast \& Quirk 1970; Hohl 1971; Ostriker \& Peebles 1973;
 Sellwood 1981; Athanassoula \& Sellwood 1986; etc.). The second
 mechanism, also confirmed in numerous $N$-body simulations, is the
 formation of tidally induced bars, triggered by interactions with
 neighbouring galaxies (e.g., Byrd et al. 1986; Noguchi 1987, 1988,
 1996; Gerin, Combes \& Athanassoula 1990; Barnes \& Hernquist 1991;
 Salo 1991; Miwa \& Noguchi 1998; etc.). To distinguish between the
 two bar forming mechanisms, Miwa \& Noguchi \shortcite{mn98}
 compared the properties of spontaneously formed and tidally induced
 bars by means of $N$-body simulations and argued that tidally
 induced bars are slow rotators, while the spontaneously formed bars
 have usually higher pattern speeds. 
 
 Athanassoula \shortcite{ath03} argued that the evolution of bars in
 isolated disc galaxies is driven by the redistribution of angular
 momentum. Since this is crucial for the work presented here, we will
 briefly summarise the main results. Disc galaxies strive to transfer
 angular momentum outwards \cite{lbk72}. Disc material in the inner
 disc, being at resonance with the bar, emits angular momentum to
 resonant material in the outer disc, or in the spheroidal components
 like the halo and the bulge \cite{ath02}. The corotation radius of
 the bar roughly divides the regions of disc emitters from that of
 disc absorbers. The size and the pattern speed of the bar responds
 to the angular momentum exchange in a way as to keep an equilibrium
 between the emitters and absorbers. These recent results argue
 strongly that the strength and the pattern speed of the bar are
 determined by the amount of angular momentum exchanged. The bar's
 slowdown rate found in numerical simulations depends on the
 relative halo mass and on the velocity dispersion of both the disc
 and the halo. Furthermore, Athanassoula \shortcite{ath02} has shown
 that bar growth in the disc can be stimulated even in massive halos
 due to the destabilising influence of resonant stars in the halo.
 
 Another factor influencing the evolution of bars is the interstellar
 medium in the disc. Owing to the gravitational torques of the bar,
 gas can be driven towards the centre of the galaxy and this is
 likely to be accompanied by central starbursts and
 formation/fueling of an active galactic nucleus (Shlosman, Frank \&
 Begelman 1989). Fully self-consistent numerical simulations
 including gas have shown that substantial gas inflow can weaken
 significantly, or even destroy, the bar (Friedli \& Benz 1993;
 Berentzen et al. 1998). The central mass concentrations and the
 times-scales found in these simulations, however, might be too high
 and too short, respectively, to be in agreement with the relatively
 high fraction of barred galaxies observed. Furthermore, this
 fraction seems to be independent of galaxy morphology, i.e. the same
 in early- and late-type disc galaxies, while early-types are known
 to have considerably less gas.

 A regeneration of the stellar bar, i.e. a secondary episode of
 bar-formation during the lifetime of the disc, has been suggested
 as a possible mechanism to explain the observed number of barred
 galaxies. One scenario for this (Sellwood \& Moore 1999; Bournaud 
 \& Combes 2002) could be the accretion of gas-rich companion
 galaxies or freshly infalling gas, which, by adding colder material 
 to the disc, may cool it sufficiently and allow the generation of a
 new bar. Another scenario suggested, and upon which we will focus in
 this paper, could be the regeneration of a previously dissolved or
 weakened bar by a tidal encounter with a neighbouring galaxy. In
 this work we study this regeneration process by means of fully
 self-consistent, 3D numerical simulations. The host galaxy, in which
 the first, i.e. spontaneously formed, bar has
 significantly weakened before the interaction, is tidally perturbed
 by a companion galaxy. Various orbits and masses have been
 considered for the companion, in order to cover a wide parameter
 space. We provide a comparison between pure stellar and
 two-component stars+gas models and describe the regeneration process
 and the dynamical properties of the tidally induced bars.
 
 In \S\ref{initial} we describe the numerical methods and the initial
 conditions of the galaxy models. The evolution and the dynamical
 properties of the isolated models with and without gas are described
 in \S\ref{hostI0} and \S\ref{hostI1}, respectively. In
 \S\ref{interI0} and \S\ref{interI1} we present the results of the
 interacting models with and without gas, respectively. The results
 are then discussed in \S\ref{discussion} and finally we give a
 summary in \S\ref{summary}.

\section{Numerical methods and initial conditions} \label{initial}

\subsection{Methods}

 To evolve both the collisionless component, representing the stars
 and the dark matter, and the dissipative component, representing the
 gas, we use an $N$-body algorithm, combined with a smoothed particle
 hydrodynamics (hereafter SPH) algorithm (e.g., review by Monaghan
 1992). For this purpose we use the version of the hybrid
 $N$-body/SPH code of Heller (1991; see also Heller \& Shlosman
 1994). The fully self-consistent 3D algorithm employs such features
 as a spatially varying smoothing length, a hierarchy of time-bins to
 approximate individual time-steps, a viscosity ``switch'' to reduce
 the effects of viscous shear, and the special-purpose hardware
 GRAPE-3AF to compute the gravitational forces and neighbour
 interaction lists (Sugimoto et al. 1990; Steinmetz 1996).

\subsection{Initial conditions} \label{initial2}

\begin{table}
\caption{Initial model parameters.}
\begin{tabular}{lccrlccc} \hline
 Component & Type &  $N_{\rm d}$  & M$_{\rm d}$ & $a_{\rm d}$ &
 r$_{\rm cut}$ & z$_0$\\ \hline
 Disc -- I0   &    &         &      &        &      &            \\
\hspace{1ex} stars & KT & 13\,500 & 0.54 &  1.0   &  5.0 & 0.20  \\
\hspace{1ex} gas   & KT & 10\,000 & 0.16 &  1.0   &  5.0 & 0.05  \\
 Disc -- I1   &    &         &      &        &      &            \\
\hspace{1ex} stars & $\cdots$ & 17\,500 & 0.7 &  $\cdots$ & $\cdots$
 & $\cdots$  \\[2ex]
\hline
 Component & Type & $N_{\rm h}$  &  M$_{\rm h}$   & $b_{\rm h}$ &
 r$_{\rm cut}$ & \\ \hline
 Halo &      &    &      &        &      &          \\
\hspace{1ex} stars & Pl & 32\,500 & 1.30 & 5.0 & 10.0 & $\cdots$
\\[2ex]
\hline
 Component & Type & $N_{\rm c}$ & M$_{\rm c}$ & &  & \\ \hline
 Companion   &  &  &  &   &  &  \\
\hspace{1ex} C1  & pt & 1 & 2.0  & $\cdots$ & $\cdots$ &
   $\cdots$  \\
\hspace{1ex} C2  & pt & 1 & 4.0  & $\cdots$ & $\cdots$ &
   $\cdots$  \\
\hspace{1ex} C3  & pt & 1 & 6.0  & $\cdots$ & $\cdots$ &
   $\cdots$  \\
\hspace{1ex} C4  & pt & 1 & 8.0  & $\cdots$ & $\cdots$ &
   $\cdots$  \\
\hspace{1ex} C5  & pt & 1 & 0.66 & $\cdots$ & $\cdots$ &
   $\cdots$  \\
\hspace{1ex} C6  & pt & 1 & 1.0  & $\cdots$ & $\cdots$ &
   $\cdots$  \\
\hline
\end{tabular}
\label{tab01}
\end{table}

 Our isolated model I0 is initially composed of a stellar and a
 gaseous disc, embedded in a spherical halo. The radial surface
 density of both the discs follows a truncated Kuzmin-Toomre
 (hereafter KT) profile (Kuzmin 1956; Toomre 1963) with a radial
 scalelength $a_{\rm d}$. Their vertical density profile follows the
 ${\rm sech}^2$ distribution of an isothermal sheet \cite{spi42}.
 The halo is set up initially as a Plummer (hereafter Pl) sphere 
 \cite{plu11} with a radial scalelength $b_{\rm h}$, and is then
 allowed to relax in the gradually introduced potential of the
 {\em frozen} disc. For a detailed description of the initial model
 I0 see also Berentzen et al. \shortcite{bahf03}. The initial model
 parameters are summarised in Tab.~\ref{tab01}. The columns from left
 to right give the component, the type of density profile, the
 particle number, the total mass, the radial scalelengths $a_{\rm d}$
 and $b_{\rm h}$, the cut-off radius $r_{\rm cut}$, and the vertical
 scaleheight $z_0$ of each component. Our isolated model I1 differs
 from model I0 in that about one third of the gas particles have been
 replaced with stellar particles, i.e. with keeping their positions
 and assigning new masses, keeping the total disc mass the same as in
 model I0. The remaining gas particles have been completely removed
 from the disc. We then assign velocities to the new stellar
 particles according to the mean velocity distribution of the
 {\em old} underlying stellar disc.  Both host galaxies, I0 and I1,
 are evolved in isolation first. The disc of I0 is constructed so as
 to be globally unstable to non-axisymmetric perturbations and form
 a large-scale bar. Owing to induced gas-inflow, the stellar bar
 in model I0 weakens significantly during its evolution, but is
 present in both models before the companion galaxy is introduced.

\begin{figure*}
 \includegraphics[width=\textwidth]{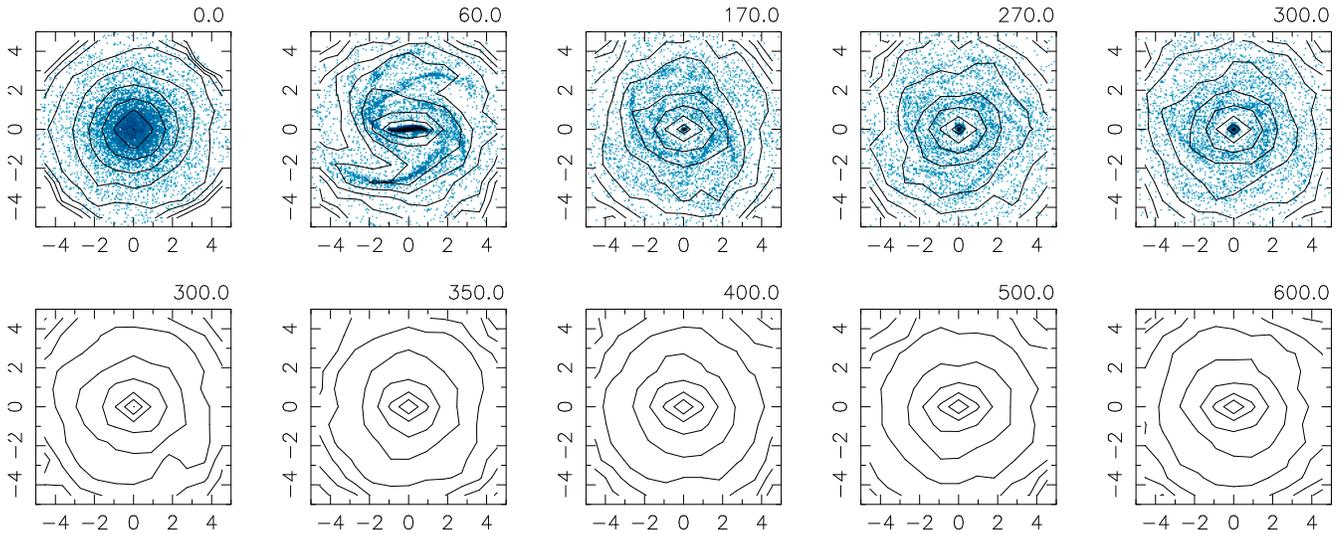}
 \caption{Evolution of the isolated models. We show the face-on
    isodensity contours of the stellar disc for models I0 (upper
    panels) and I1 (lower panels) and a grey-scale density plot of
    the gas particle distribution for model I0. The disc rotation
    is counter-clockwise, and the disc has been rotated in each frame
    so that the bar is oriented along the $x$-axis. We give the
    time on the top of each panel. Time and length scales are
    both in computer units.}
 \label{fig01}
\end{figure*}

 The companion galaxy in the interaction simulations is represented
 by a smoothed point mass (hereafter pt) for simplicity, applying a
 Plummer softening for the force calculation. Since the GRAPE-3
 hardware can only handle a fixed softening length, we apply the same
 softening for all particles, including the companion. The mass ratio
 $M_{\rm host}\!:\!M_{\rm comp}$, where $M_{\rm host}$ and
 $M_{\rm comp}$ are the total mass of the host and of the companion
 galaxy, respectively, has been varied between $1\!:\!1$ to
 $1\!:\!4$. In addition a few simulations have been run with less
 massive companions. For the simulations presented in this paper
 we confine ourselves to planar fly-by encounters, i.e. the orbits of
 the host and the companion galaxy are unbound and lie in the
 equatorial plane of the host. The initial conditions have been
 chosen such as to provide a strong perturbation to the disc. The
 orbit of the companion galaxy is prograde with respect to the
 rotation in the host disc and has been chosen to be either parabolic
 or hyperbolic, with different pericentric separations
 $R_{\rm peri}$. The pericentric position has been chosen such that
 at time of pericentre (hereafter $t_{\rm peri}$) the major axis of
 the relatively weak bar, which has formed/evolved in the hosts disc
 before the encounter, points towards the companion. The times of
 pericentric separation have been chosen to be $t_{\rm peri}\!=\!270$
 and $500$ for all simulations with host galaxy I0 and I1,
 respectively. The initial orbital parameters at $t_{\rm peri}$ have
 been calculated from the solution of the corresponding two-body
 problem, either for a given orbital eccentricity $e$, or for a
 pericentric angular frequency of the companion
\begin{equation}
  \omega_{\rm peri} =  v_{\rm peri}/R_{\rm peri} \ ,
\end{equation}
 where $v_{\rm peri}$ is the velocity of the companion at pericentre.
 To obtain the initial centre of mass positions and velocities of the
 host and of the companion galaxy, we integrate their orbits backward
 in time, starting from the pericentric time $t_{\rm peri}$, until
 the distance $\Delta R$ between host and companion is roughly 10
 times the cut-off radius of the initial halo. During this backwards
 orbit integration the particles of the host are frozen with respect
 to each other. Owing to the limited spatial range of the GRAPE-3
 hardware, the interaction models have been calculated in the
 inertial frame of the host galaxy and the force of the companion has
 partly been calculated by direct summation on the front-end, i.e.
 without using the GRAPE hardware.

\subsection{Units}

 The adopted units for mass, distance, and time are
 $M\!= 6\times10^{10}$\,M$_\odot$, $\rm{R}\!=\!3$\,kpc and
 $\tau\!=\!10^7$ yr, respectively, for which the gravitational
 constant G is unity. The dynamical time is $\tau_{\rm dyn}
 \!\equiv\!(r^3_{\rm 1\!/\!2}/{\rm G}\,M_{\rm 1\!/\!2})^{1/2}
 \!=\!4.8\times10^7$ yr, where $M_{\rm 1\!/\!2}$ is half the total
 mass of the host and $r_{1\!/\!2}$ is its half-mass radius, which
 is, after relaxation of the halo, approximately 8.5\,kpc or in
 terms of the disc scalelength, $2.8\,a_{\rm d}$. The initial
 stellar disc rotation period in these units then corresponds to
 $t_{\rm rot}\!\equiv\!2\pi\tau_{\rm dyn} \approx 3\!\times\!
 10^8$\,yr.  A fixed gravitational softening length of
 $\epsilon\!=\!0.375$\,kpc is used for all particles. An isothermal
 equation of state is used for the gas with a sound speed of
 $v_{\rm s}\!=\!12$\,km s$^{-1}$.  The corresponding thermal gas
 temperature is 10$^4$\,K.

\section{Isolated model with gas -- model I0} \label{hostI0}

\begin{figure}
\begin{center}
 \includegraphics[angle=-90,width=\columnwidth]{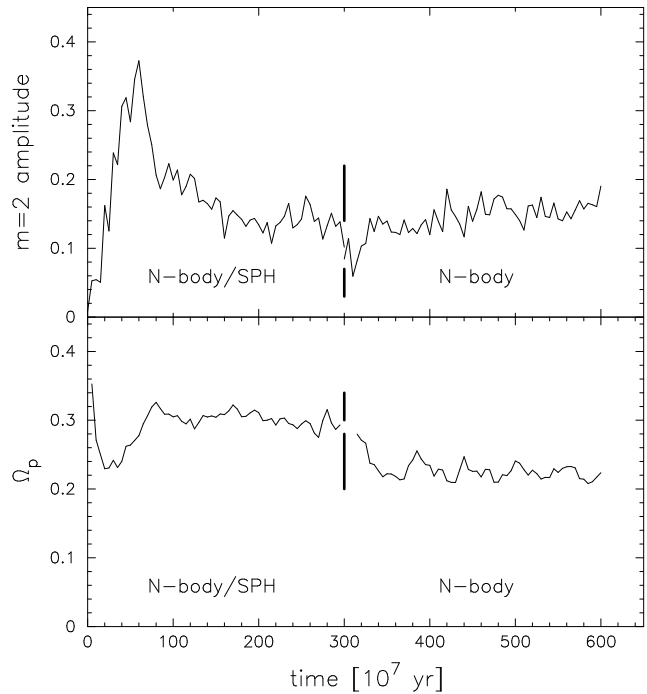}
\end{center}
 \caption{Evolution of the bar strength (upper panel) and pattern
   speed (lower panel) for models I0 (stars+gas) and
   I1 (stars). The transition between the two models
   is marked by two vertical dashes ($t\!=\!300$). }
 \label{fig02}
\end{figure}

 As described in the previous section, the isolated galaxy model I0
 is composed of a stellar (collisionless) and a gaseous (dissipative)
 disc component, embedded in a {\em live} halo. The morphological
 evolution of the disc is shown in Fig.~\ref{fig01} (upper panels).
 The stellar disc has been chosen to be initially bar unstable and
 forms a large-scale bar within roughly 2\,$t_{\rm rot}$. The stellar
 bar reaches its maximum strength at about $t\!=\!60$ (see
 Fig.~\ref{fig02}). The bar strength, here defined by the normalized
 $m\!=\!2$ Fourier component of the stellar disc mass distribution,
 has been measured inside a cylindrical radius of $R\!=\!1.25$ and
 about one scaleheight of the disc plane. Owing to the torque of the
 bar, most of the gas in the bar region is driven towards the centre
 of the galaxy, accumulating in a nuclear and circumnuclear disc. At
 the end of the run the mass of the gaseous nuclear disc represents
 some 18 per cent of the total mass within a spherical radius of
 $1.0$\,kpc. As a result of the growing central mass concentration
 the stellar bar weakens significantly, and the disc settles down in
 a quasi-stable state at about $t\!\approx\!120$. Some gas also
 accumulates  near the end of the weak stellar bar, forming an oval
 ring close to the position of the inner ultra-harmonic resonance
 (hereafter UHR) at $R\!=\!2.25$\,kpc. The pattern speed of the bar
 (hereafter $\Omega_{\rm p}$) first increases due to the short burst
 of gas inflow and thereafter reaches a constant rate of about
 $\Omega_{\rm p}\!=\!0.3\,\tau^{-1}$, or 29.3 km s$^{-1}$ kpc$^{-1}$.
 The evolution of model I0 is followed in total up to $t\!=\!300$,
 i.e. about 10 disc rotations. A more detailed description of the
 evolution of this model and its dynamical properties is given in
 Berentzen et al. \shortcite{bahf03}.

 To identify the presence and the location of the main planar
 resonances in the disc, we apply both the linear (epicyclic)
 approximation and non-linear methods. We therefore calculate the
 gravitational potential in the $z\!=\!0$ plane of the disc on a
 Cartesian grid and symmetrise it with a four-fold symmetry with
 respect to the main axes of the bar. From the potential we then
 derive the azimuthally averaged circular and epicyclic frequency
 $\Omega_{\rm c}(R)$ and $\kappa(R)$, respectively. The standard
 linear resonance condition is given by 
\begin{equation}
 \Omega_{\rm p} = \Omega_{\rm c} + \frac{l}{m}\,\kappa \ ,
\end{equation}
 with integers $l$ and $m$. For a strict definition of the inner
 Lindblad resonances (hereafter ILRs; with $l\!=\!-1$ and $m\!=\!2$)
 three extensions to the linear definition have been proposed 
 (see Athanassoula \shortcite{ath03} for a discussion), which are not
 fully compatible to each other and 
 sometimes in contradiction. Throughout this work, unless otherwise
 noted, we apply the orbital structure definition for the ILRs, i.e.
 by saying that an
 ILR exists if and only if the $x_2$ and $x_3$ orbit families exist. 
 The ILRs have been confirmed in this case by calculating the
 surfaces of section (hereafter SOS). The SOS have been constructed
 by integrating orbits of a given Jacobian energy (hereafter
 E$_{\rm J}$; see also Sec.~\ref{hostI1}) in the equatorial plane of
 the disc and marking the points in the $(y,\dot{y})$ plane each time
 the orbits cross the line $x\!=\!0$ with $\dot{x}<0$ (e.g., Binney
 \& Tremaine 1987).  At the end of the run ($t\!=\!300$) the
 corotation radius (hereafter CR; i.e., the radius at which
 $\Omega_{\rm c}\!=\!\Omega_{\rm p}$) in this model is at about
 $6.15$\,kpc ($E_{\rm J}\!\approx\!-0.91$), and the ILRs are located at
 about 0.47\,kpc (inner ILR; $E_{\rm J}\!=\!-1.58$) and 2.1\,kpc
 (outer ILR; $E_{\rm J}\!=\!-1.10$), respectively.  

\begin{figure} 
\begin{center}
 \includegraphics[angle=-90, width=\columnwidth]{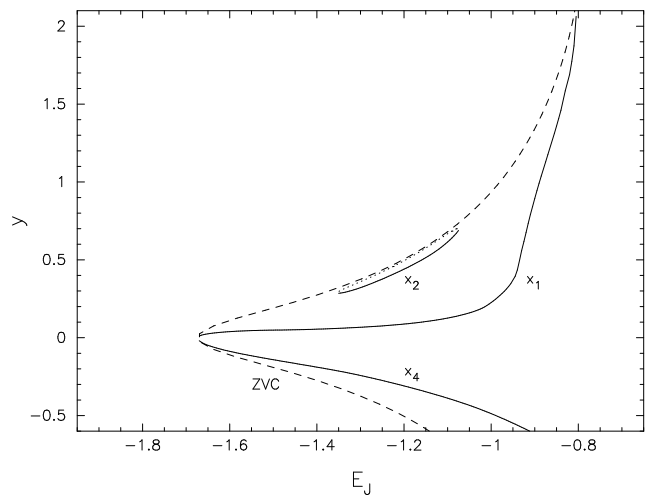}
\end{center}
 \caption{Characteristic diagram of the isolated model I1 at the 
  end of the run. We show the characteristic curves of the main
  orbit families (full lines) and the zero velocity curve (ZVC;
  dashed line).}
 \label{fig03}
\end{figure}

\section{Isolated model without gas -- model I1} \label{hostI1}

\begin{figure*} 
 \includegraphics[width=\textwidth]{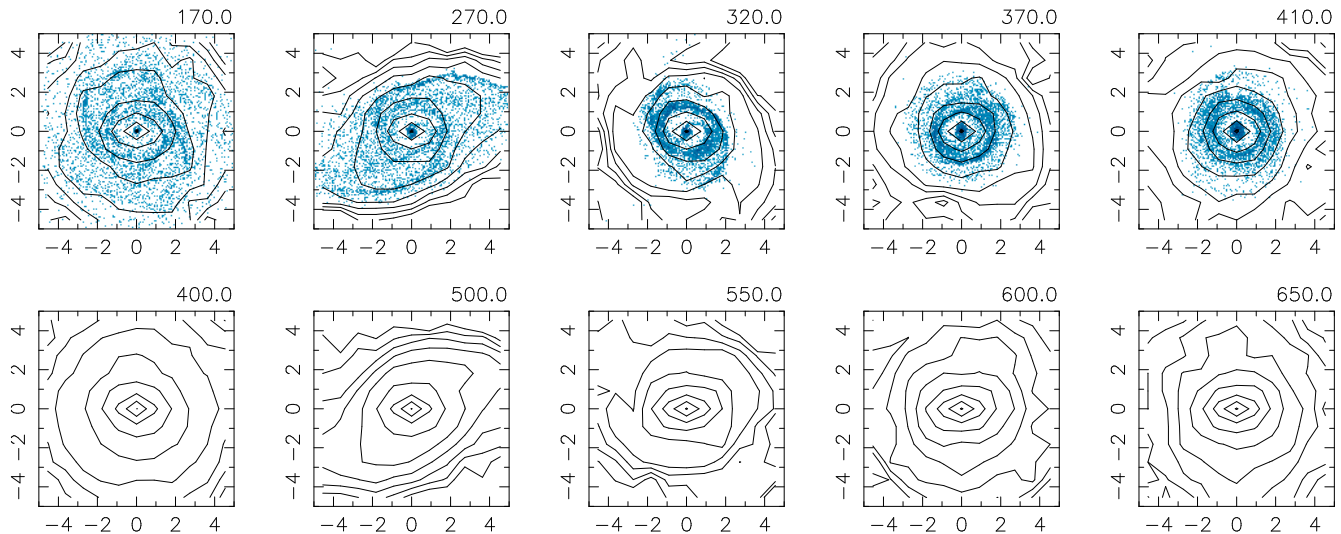}
 \caption{Examples of the morphological evolution of interaction
    models with and without gas. We show the face-on isodensity
    contours of the stellar disc for the models I0\,C4\,p (with gas,
    upper panels) and I1\,C4\,p (without gas, lower panels). For the
    first model a grey-scale plot of the gas particle distribution is
    shown. The layout is as for Fig.~\ref{fig01}.}
 \label{fig04}
\end{figure*}

 We study the regeneration of the bar by interactions of two
 different types of host galaxies -- with and without gas in the disc
 -- which, nevertheless, have similar dynamical properties. We
 therefore construct our purely stellar model I1 by replacing the gas
 particles in model I0 with stellar particles, as described in
 Sec.~\ref{initial2}. This has been accomplished for the last
 snapshot of model I0 at $t\!=\!300$, i.e. at the end of the run,
 when the amplitude of the initial bar has very considerably decreased
 and the disc is in a 
 quasi-stable state. After replacing the gaseous particles with
 stellar ones we follow the evolution of the new model I1 for another
 $\Delta t\!=\!300$. The morphological evolution of the stellar disc
 is shown in Fig.~\ref{fig01} (lower panels), and in Fig.~\ref{fig02}
 we show the evolution of the bar strength and of the pattern speed.
 The bar strength first decreases abruptly, owing to the {\em newly}
 added stellar particles, which follow initially the distribution of
 the gas. The phase angle of the $m\!=\!2$ Fourier component of the
 latter is shifted with respect to that of the old
 underlying disc and therefore tends to decrease the normalized
 $m\!=\!2$ amplitude of the combined discs. The velocities, which are
 assigned to the new stellar disc component, follow the velocity
 distribution of the {\em old} underlying stellar disc. Therefore the
 stellar disc of model I1 is initially slightly out of virial
 equilibrium, since the dispersion of the stellar particles is higher
 than that of the gas particles. After some $\Delta t\!=\!30$,
 however, the disc settles to a new equilibrium and the bar strength
 increases again, reaching its previous value and remaining constant
 after the short relaxation 
 phase. Owing to the decreasing central mass concentration of the
 new stellar component, the pattern speed of the bar slightly
 decreases first, but remains roughly constant after the relaxation
 phase at about $\Omega_{\rm p}\!=\!0.22 \tau^{-1}$, or
 21.5 \,km s$^{-1}$\,kpc$^{-1}$, till the end of the run.

\begin{figure} 
\begin{center}
\includegraphics[angle=-90,width=\columnwidth]{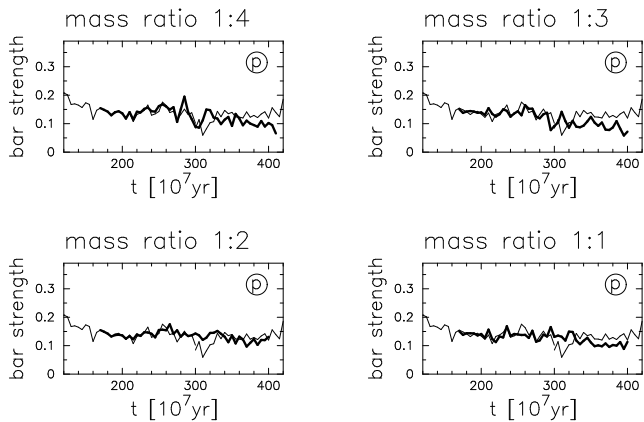}
\end{center}
 \caption{Evolution of the bar strength as a function of time for
  prograde, parabolic encounters with host galaxy I0
  (thick line). The pericentric separation is $R_{\rm peri}\!=\!15$
  and the mass ratio $M_{\rm host}\!:\!M_{\rm comp}$ is given on top
  of each frame. The type of the companions orbit is given in the 
  upper right corner of each frame. For comparison we also show
  the bar strength of the isolated model (thin line).}
 \label{fig05}
\end{figure}

\begin{figure} 
\begin{center}
\includegraphics[angle=-90,width=\columnwidth]{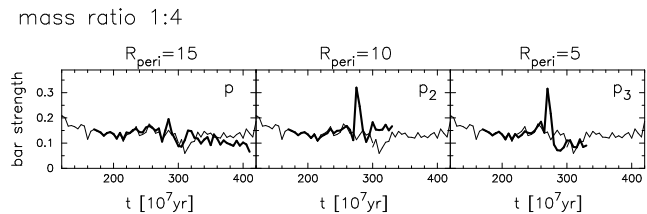}
\end{center}
 \caption{Same as Fig.~\ref{fig05}, but for encounters with
  a fixed mass ratio of $1\!:\!4$ and pericentric separations
  of 15 (p), 10 (p$_2$) and 5 (p$_3$), respectively.}
 \label{fig06}
\end{figure}

 To gain some insight in the orbital structure of the disc of model
 I1, we locate the main planar periodic orbits in a frozen potential
 that rotates with the bar. General information on orbits in barred
 potentials can be found, e.g., in Contopoulos \& Grosb{\o}l
 \shortcite{cg89}. The gravitational potential has been calculated
 the same way as described in the previous section, but is
 time-averaged over roughly one bar rotation. For simplicity, we
 restrict ourselves to planar orbits which are bi-symmetric with
 respect to the bar and close after one orbit around the centre in
 the frame of reference corotating with the bar. The results are
 then displayed in terms of a characteristic diagram, where for each
 orbit we plot its Jacobi integral $E_{\rm J}$ with respect to the
 $y$-intercept value with the $x\!=\!0$ plane. The Jacobi integral
 is a conserved quantity along any given orbit in the rotating frame,
 and can be thought of as an effective energy (e.g., Binney \&
 Tremaine 1987). In the characteristic diagram the orbits form curves
 of families.  In Fig.~\ref{fig03} we show the characteristic diagram
 of the main orbit families, using the notation of Contopoulos \&
 Papayannopoulos \shortcite{cp80}, in model I1 at the end of the
 run ($t\!=\!600$).  The dashed curve is the zero-velocity curve
 (hereafter ZVC), which delineates the accessible region in the plane
 based on energy considerations. The family labelled $x_1$ consists
 of orbits that are elongated along the bar and predominately gives
 the bar its structure. Orbits of the $x_2$-family are elongated
 perpendicular to the bar. Their presence is indicative of an ILR
 (or more than one) in the non-linear regime. The $x_4$ orbits are
 retrograde and slightly elongated perpendicular to the bar. The
 corotation radius in model I1 is at $E_{\rm J}\!=\!-0.79$, or
 approximately $7.98$\,kpc. The $x_2$-orbits range in energy from
 $E_{\rm J}\!=\!-1.35$ to $-1.07$, with semi-major axes ranging from
 0.86 to 2.1\,kpc.

\section{Interacting models with gas} \label{interI0}

\begin{table}
 \caption{Notation of the interaction models.}
 \begin{center}
 \begin{tabular}{c|c|c|c|c|} \hline
           &     \multicolumn{3}{c}{$\omega_{\rm peri}$} & para- \\
 $R_{\rm peri}$  &  0.3 & 0.6 & 0.9 & bolic \\ \hline 
  15.0 &    a &  b &  c & p      \\ 
  10.0 &    d &  e &  f & p$_1$  \\
   5.0 &    g &  h &  i & p$_2$  \\ \hline
 \end{tabular}
 \end{center}
\label{tab02}
\end{table}

 In this section we describe the simulations in which the gas-rich
 host galaxy I0 is perturbed by encounters with companions of
 different mass. The time of pericentric separation has been chosen
 to be $t_{\rm peri}\!=\!270$ for all runs with the host galaxy I0.
 In the first set of simulations the companions are initially set up
 to follow prograde, parabolic orbits with a pericentric separation
 of $R_{\rm peri}\!=\!15$, which is just outside the halo of the host
 galaxy. When referring to a specific model, we will hereafter use a
 notation like, for instance, `I0\,C4\,p', where `I0' is the name of
 the host galaxy, `C4' is the name of the companion (see
 Tab.~\ref{tab01}), and `p' (parabolic) denotes the type of orbit,
 the latter following the notation given in Tab.~\ref{tab02}. In
 Tab.~\ref{tab03} we summarise the basic parameters and properties of
 both the isolated and the interaction models. The first and the
 second column give the name of the host and of the companion galaxy,
 respectively, and the third column gives the orbit of the companion
 (see Tab.~\ref{tab02}). In the fourth column we give the time when
 the companion is included in the simulation, resulting from the
 initial backward orbit integration (see Sec.~\ref{initial2}). The fifth to
 seventh column give the pericentric separation and frequency, and
 the eccentricity of the orbit, respectively. In the eighth and ninth
 column we give the strength and pattern speed of the bar,
 respectively, after the interaction. The tenth and eleventh columns
 give the ratio of bar strength and pattern speed to the
 corresponding values of the isolated model. In the penultimate
 column we give the interaction strength parameter as defined in 
 Sec.~\ref{interI1}.1 and in the last column the corotation radius
 at the end of each run, determined from the linear analysis.

 In Fig.~\ref{fig04} (upper panels) we show an example of the
 morphological evolution of the disc during an encounter, in this
 case for model I0\,C4\,p. The bar strength as a function of time for
 the models with different massive companions is shown in
 Fig.~\ref{fig05}. We find that the encounters in this set of
 simulations are not sufficiently strong to regenerate the stellar
 bar in the disc, even for a mass ratio of $1\!:\!4$, i.e. even with
 companion C4. In order to increase the strength of the tidal
 perturbation on the host galaxy, we run two additional simulations
 with companion C4, in which we now choose pericentric
 separations of $R_{\rm peri}\!=\!10$ and $5$. The orbits of the
 companion are again chosen to be prograde and parabolic. The bar
 strength for these models is shown in Fig.~\ref{fig06}. The peaks
 immediately following $t_{\rm peri}$ result from the time when the
 transient tidal arms contribute to the $m\!=\!2$ power in the inner
 disc region, in which the bar amplitude is measured, but disappear
 as soon as the tidal arm features have dissolved. Even with these
 close encounters, it is not possible to regenerate the stellar
 bar in the disc of the gas-rich host galaxy.

\begin{figure}
\begin{center}
\includegraphics[angle=-90, width=\columnwidth]{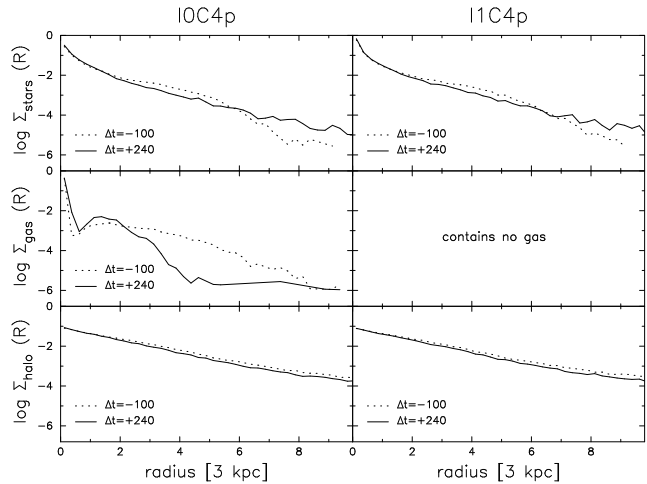}
\end{center}
 \caption{Logarithmic surface density as a function of radius
   for the stellar and gaseous disc and of the halo (from top
   to bottom, respectively). We show ${\rm log} \Sigma(R)$
   for models I0\,C4\,p (left-hand panels) and I1\,C4\,p
   (right-hand panels). The dotted and full lines
   represent times before ($\Delta t\!=\!-100$) and after
   ($\Delta t\!=\!240$) the interaction, respectively.}
 \label{fig07}
\end{figure}

\begin{figure}
 \begin{center}
 \includegraphics[angle=-90, width=\columnwidth]{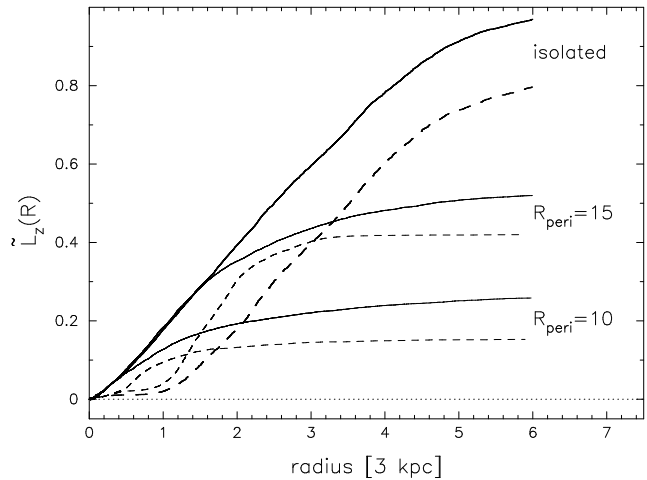}
 \end{center}
 \caption{Specific cumulative angular momentum ${\tilde L}_z$ as a
    function of radius in the stellar disc at the end of the run of
    the isolated model I0 (with gas) and of the interaction models
    with a mass ratio of $1\!:\!4$ and different pericentric
    separations. The full and dashed lines show the results for
    the stellar and gaseous disc, respectively.}
 \label{fig08}
\end{figure}

\begin{figure}
 \begin{center}
 \includegraphics[angle=-90, width=\columnwidth]{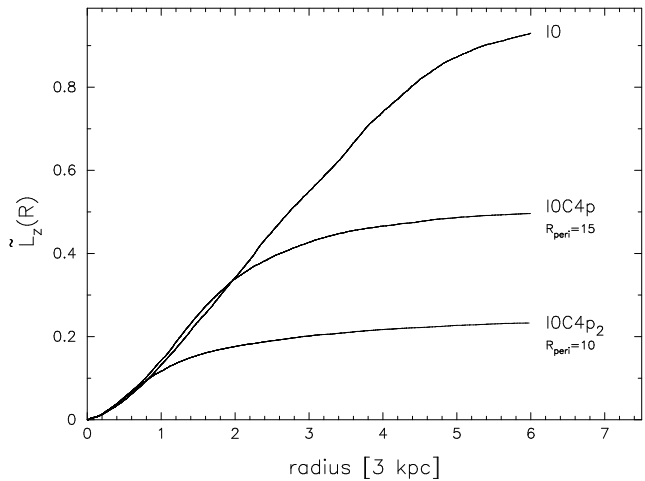}
 \end{center}
 \caption{Same as Fig.~\ref{fig08}, but for the
   combined disc (stars+gas).}
 \label{fig09}
\end{figure}

 In order to facilitate comparison with the purely stellar simulations
 described in the next section, we present here a more detailed
 analysis of the interaction simulations with the gas-rich host
 galaxy I0. As can be seen in Fig.~\ref{fig04} already, the
 interaction apparently leads to a significant redistribution of both
 the stellar and the gaseous material in the disc. This becomes more
 evident in Fig.~\ref{fig07}, in which we show the logarithmic radial
 surface density of the disc and the halo for model I0\,C4\,p at
 times before and after the interaction. While the stellar density
 remains basically constant within a radius of approximately
 $R\!<\!2.0$, i.e. roughly inside corotation in the corresponding
 isolated model, there is some net inflow of gas from the outer disc.
 The inflowing gas first piles up at a radius of about
 $R\!\approx\!1.25$, which is close to the position of the gaseous
 ring, and then flows gradually towards the central disc region,
 accumulating in the nuclear disc. At the end of the run, the gas
 mass within 0.5 length units has increased by about 0.5 per cent of
 the total host mass, or some 6 per cent of the total gas mass, while
 the gas mass within 2 length units increased by roughly 1 per cent
 of the total mass, or about 12 per cent of the total gas mass. Some
 fraction of both the stellar and the gaseous material in the outer
 disc moves further out, contributing to the transient tidal arms
 and/or getting stripped off by the companion. The halo material in
 the inner region does not show any net radial redistribution, but we
 find some expansion in the outer parts, where halo material is also
 stripped off by the companion.  

\begin{table*}
\caption{Main model parameters and properties.}
\begin{tabular}{lcccccccccccc}
 \hline
 Host & Comp  & orbit & t$_{start}$ & $R_{\rm peri}$ & $\omega_{\rm
peri} $ & $e$ & A$_{(m\!=\!2)}$ & $\Omega_{\rm p}$ &
 A/A$_{\rm I0,I1}$ & $\Omega_{\rm p}/\Omega_{\rm I0,I1}$ & $\Theta$ &
 CR \\
\hline
I0 &  --   &  --   &   0.0 & $\cdots$ & $\cdots$ & $\cdots$ & 0.14  & 0.30  &   1.0 &  1.0 & $\cdots$ & 2.05 \\
I1 &  --   &  --   & 100.0 & $\cdots$ & $\cdots$ & $\cdots$ & 0.16  & 0.22  &   1.0 &  1.0 & $\cdots$ & 2.76 \\
 \hline
I0 & C1 & p & 170.0 & 15.0     & 0.049   &   1.00   & 0.10 & 0.30 & 0.73 & 1.33 & 1.927 & 1.80 \\
I0 & C2 & p & 170.0 & 15.0     & 0.060   &   1.00   & 0.13 & 0.30 & 0.91 & 1.01 & 3.159 & 2.13 \\
I0 & C3 & p & 170.0 & 15.0     & 0.069   &   1.00   & 0.08 & 0.27 & 0.61 & 0.92 & 4.116 & 2.31 \\
I0 & C4 & p & 170.0 & 15.0     & 0.077   &   1.00   & 0.11 & 0.26 & 0.77 & 0.87 & 4.922 & 2.47 \\
   &  & p$_2$ & 170.0 & 10.0     & 0.141   &   1.00   & 0.16 & 0.18 & 1.19 & 0.63 &       & 2.99 \\
   &  & p$_3$ & 170.0 &  5.0     & 0.400   &   1.00   & 0.10 & 0.22 & 0.70 & 0.74 &       & $\cdots$ \\
I0 & C4 & d & 250.0 & 10.0     & 0.300   &   8.00   & 0.06 & 0.28 & 0.44 & 0.93 & 1.906 & 2.20 \\
 \hline
I1 & C5 & g & 450.0 &  5.0     & 0.300   &   3.22   & 0.18 & 0.20 & 1.14 & 0.90 & 0.717 & 2.94 \\
 \hline
I1 & C6 & j & 470.0 &  7.5     & 0.300   &  11.66   & 0.17 & 0.21 & 1.08 & 0.92 & 0.419 & 2.94 \\
I1 & C6 & g & 450.0 &  5.0     & 0.300   &   2.75   & 0.21 & 0.19 & 1.31 & 0.82 & 1.105 & 2.78 \\
 \hline
I1 & C1 & p & 400.0 & 15.0     & 0.049   &   1.00   & 0.16 & 0.22 & 1.00 & 0.98 & 1.927 & 2.87 \\
I1 & C1 & d & 480.0 & 10.0     & 0.300   &  21.50   & 0.17 & 0.21 & 1.11 & 0.93 & 0.458 & 2.95 \\
I1 & C1 & e & 490.0 & 10.0     & 0.600   &  89.00   & 0.16 & 0.21 & 1.04 & 0.95 & 0.261 & 2.90 \\
I1 & C1 & g & 440.0 &  5.0     & 0.300   &   1.81   & 0.27 & 0.12 & 1.74 & 0.56 & 2.417 & 4.13 \\
I1 & C1 & h & 480.0 &  5.0     & 0.600   &  10.25   & 0.22 & 0.18 & 1.38 & 0.82 & 1.124 & 3.25 \\
I1 & C1 & i & 480.0 &  5.0     & 0.900   &  24.31   & 0.19 & 0.19 & 1.19 & 0.87 & 0.833 & 3.20 \\
I1 & C1 & j & 470.0 &  7.5     & 0.300   &   8.48   & 0.21 & 0.19 & 1.33 & 0.84 & 0.852 & 3.19 \\
 \hline
I1 & C2 & p & 400.0 & 15.0     & 0.060   &   1.00   & 0.17 & 0.21 & 1.08 & 0.94 & 3.159 & 2.95 \\
I1 & C2 & d & 480.0 & 10.0     & 0.300   &  14.00   & 0.18 & 0.20 & 1.15 & 0.92 & 0.928 & 2.99 \\
I1 & C2 & e & 490.0 & 10.0     & 0.600   &  59.00   & 0.17 & 0.21 & 1.09 & 0.95 & 0.523 & 2.90 \\
I1 & C2 & h & 480.0 &  5.0     & 0.600   &   6.50   & 0.21 & 0.18 & 1.33 & 0.81 & 2.303 & 3.27 \\
I1 & C2 & i & 480.0 &  5.0     & 0.900   &  15.88   & 0.20 & 0.19 & 1.25 & 0.87 & 1.679 & 3.10 \\
 \hline
I1 & C3 & p & 400.0 & 15.0     & 0.069   &   1.00   & 0.21 & 0.18 & 1.36 & 0.81 & 4.116 & 3.30 \\
I1 & C3 & a & 480.0 & 15.0     & 0.300   &  36.97   & 0.17 & 0.21 & 1.05 & 0.94 & 0.500 & 2.86 \\
I1 & C3 & d & 480.0 & 10.0     & 0.300   &  10.25   & 0.21 & 0.18 & 1.36 & 0.82 & 1.410 & 3.29 \\
I1 & C3 & e & 490.0 & 10.0     & 0.600   &  44.00   & 0.19 & 0.20 & 1.18 & 0.89 & 0.787 & 3.07 \\
I1 & C3 & f & 490.0 & 10.0     & 0.900   & 100.25   & 0.17 & 0.20 & 1.09 & 0.94 & 0.594 & 2.91 \\
I1 & C3 & h & 480.0 &  5.0     & 0.600   &   4.63   & 0.26 & 0.13 & 1.64 & 0.61 & 3.545 & 3.71 \\
I1 & C3 & i & 480.0 &  5.0     & 0.900   &  11.66   & 0.24 & 0.17 & 1.51 & 0.74 & 2.541 & 3.35 \\
 \hline
I1 & C4 & p & 400.0 & 15.0     & 0.077   &   1.00   & 0.23 & 0.17 & 1.45 & 0.77 & 4.922 & 3.43 \\
I1 & C4 & a & 480.0 & 15.0     & 0.300   &  29.38   & 0.17 & 0.21 & 1.09 & 0.94 & 0.803 & 2.93 \\
I1 & C4 & b & 490.0 & 15.0     & 0.600   & 120.50   & 0.16 & 0.21 & 1.05 & 0.95 & 0.460 & 2.88 \\
I1 & C4 & c & 490.0 & 15.0     & 0.900   & 272.38   & 0.16 & 0.22 & 1.03 & 0.97 & 0.349 & 2.82 \\
I1 & C4 & d & 480.0 & 10.0     & 0.300   &   8.00   & 0.24 & 0.17 & 1.52 & 0.74 & 1.906 & 3.51 \\
I1 & C4 & e & 490.0 & 10.0     & 0.600   &  35.00   & 0.17 & 0.20 & 1.11 & 0.90 & 1.052 & 3.05 \\
I1 & C4 & f & 490.0 & 10.0     & 0.900   &  80.00   & 0.17 & 0.21 & 1.10 & 0.93 & 0.793 & 2.94 \\
I1 & C4 & h & 470.0 &  5.0     & 0.600   &   3.50   & 0.22 & 0.13 & 1.41 & 0.57 & 4.869 & 3.66 \\
I1 & C4 & i & 480.0 &  5.0     & 0.900   &   9.13   & 0.24 & 0.16 & 1.50 & 0.70 & 3.419 & 3.41 \\
 \hline
\end{tabular}
\label{tab03}
\end{table*}

\begin{figure} 
\begin{center}
\includegraphics[angle=-90, width=\columnwidth]{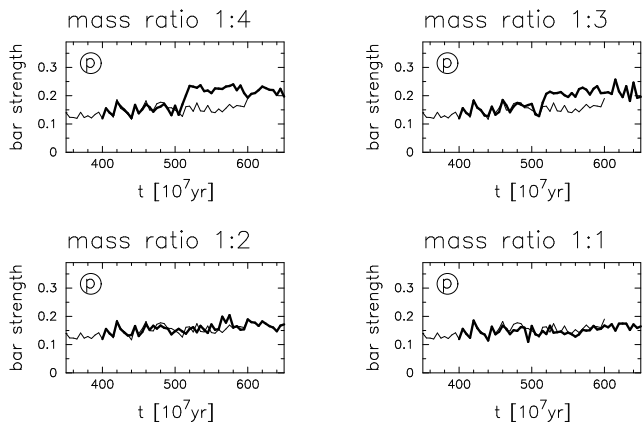}
\end{center}
 \caption{Bar strength as a function of time for the prograde,
  parabolic encounters with the purely stellar model I1 (thick line).
  The layout is as in Fig.~\ref{fig05}.}
 \label{fig10}
\end{figure}

\begin{figure}
\begin{center}
 \includegraphics[angle=-90, width=\columnwidth]{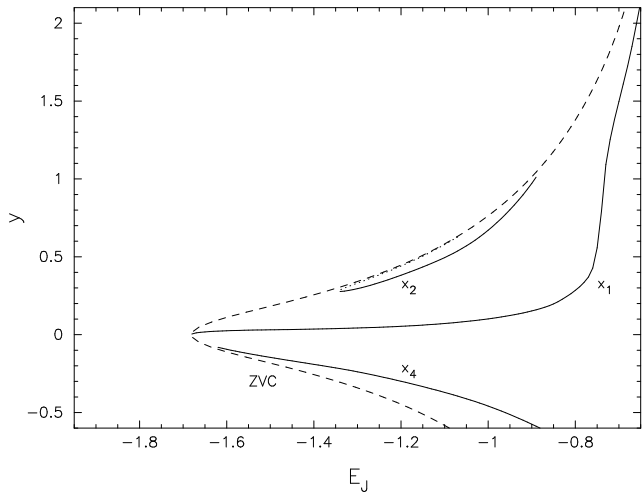}
\end{center}
 \caption{Characteristic diagram of model I1\,C4\,p after the
   interaction. The layout is as in Fig.~\ref{fig03}.}
 \label{fig11}
\end{figure}

 We measure also the specific cumulative angular momentum (hereafter
 $\tilde{L}_z$) as a function of radius for both the stellar and the
 gaseous disc, given by
\begin{equation}
 \tilde{L}_z(R) =
     \left(  \mathop{\sum_{R_i < R}} l_{z,i} \right) \times
     \left(  \mathop{\sum_{R_i < R}} m_i \right)^{-1}   \  ,
\end{equation}
 where $R_i$, $m_i$ and $l_{z,i}$ are the planar radius, the mass and
 the angular momentum, respectively, of the $i$-th disc particle. As
 an example we show in Fig.~\ref{fig08} the results of simulations
 with companion C4 and different pericentric separations. As can be
 seen in this plot, the interaction removes angular momentum from the
 stellar disc at all radii shown. We do not find any indication that 
 the corotation radius of the bar or of the companion separates
 regions which gain from regions which lose angular momentum. More
 angular momentum is removed from the stellar disc for the closer
 pericentric separation. The gaseous disc shows a different
 behaviour. The gas in the inner disc gains angular momentum compared
 to the isolated model, within roughly $R\!=\!3.0$ and $1.7$ for
 pericentric separations of $R_{\rm peri}\!=\!15$ and $10$,
 respectively, and loses angular momentum outside those radii.
  
 In Fig.~\ref{fig09} we show the specific cumulative angular momentum
 as a function of radius in the combined disc (stars+gas). We find
 that the disc in the interaction models I0\,C4\,p and I0\,C4\,p$_2$
 does not lose, or even gains, angular momentum within a radius of
 $R\!\approx\!2.0$ and $R\!\approx\!1.0$, respectively, as compared
 to the isolated model I0.

\section{Interacting models without gas} \label{interI1}

\begin{figure} 
 \begin{center}
 \includegraphics[angle=-90, width=\columnwidth]{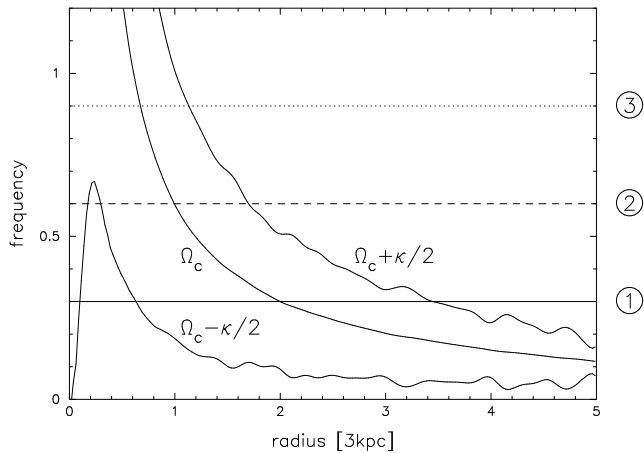}
 \end{center}
 \caption{Basic frequencies of the isolated model I1
     ($t\!=\!500$) as a function of radius, obtained with the linear
     axisymmetric definition, using the epicyclic approximation. The
     horizontal lines mark the angular frequency $\omega_{\rm peri}$
     of the companion at the time of pericentric separation.}
 \label{fig12}
\end{figure}

\begin{figure} 
\begin{center}
 \includegraphics[angle=-90, width=\columnwidth]{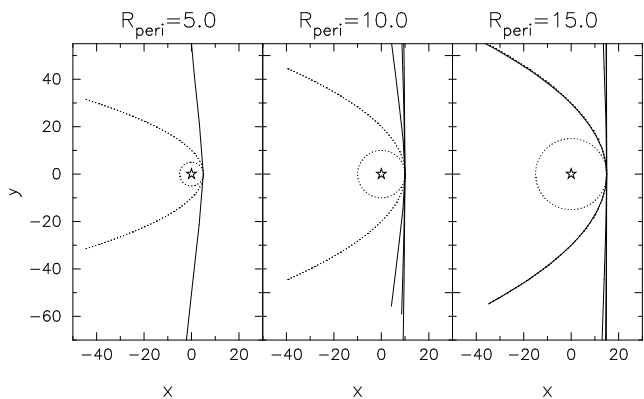}
\end{center}
 \caption{Different trajectories of the companion galaxy C4. We show
    the parabolic and hyperbolic orbits (full lines) from simulations
    with different pericentric separations, as given on top of each
    panel. The corresponding circular and parabolic orbits are
    indicated by dotted lines. The host galaxy is marked by a star
    in each panel.}
 \label{fig13}
\end{figure}

 In this section we describe the set of simulations with the purely
 stellar model I1 as the host galaxy. We start with simulations
 having similar orbits for the interaction as the ones described in
 the previous section, i.e. prograde and parabolic encounters using
 different companions (see Tab.~\ref{tab03}). As shown in
 Fig.~\ref{fig10}, in which we plot the bar strength as a function of
 time, simulations with mass ratios of $1\!:\!3$ and $1\!:\!4$ are,
 contrary to the dissipative models, sufficiently strong to
 regenerate the stellar bar in the disc, and the strength of the
 induced bar increases with the mass of the companion. In
 Fig.~\ref{fig11} we show as an example the characteristic diagram of
 the host galaxy of the interaction model I1\,C4\,p at time
 $t\!=\!640$, i.e.  the end of the run, in which the stellar bar has
 been successfully regenerated by the interaction. The orbit analysis
 has been performed following the description given in
 Sec.~\ref{hostI1}. The corotation radius in this model is at
 $E_{\rm J}\!=\!-0.63$, or about $10.3$\,kpc. The $x_2$ orbits range
 in energy from $E_{\rm J}\!=\!-1.34$ to $-0.88$, with semi-major
 axes ranging from $0.83$ to $3.04$\,kpc.

 Motivated by these results, we now explore a much wider -- though
 still far from complete -- parameter range, in order to determine
 the initial conditions necessary (or sufficient) to regenerate the
 stellar bar in the disc. We therefore run a set of simulations, in
 which we vary the orbit of the companion, the pericentric separation
 $R_{\rm peri}$ and the mass $M_{\rm C}$, respectively. The
 simulations and their basic initial parameters are summarised in
 Tab.~\ref{tab03}. From the solution of the corresponding two-body
 problem the eccentricity $e$ of the companions orbit has been
 determined by its angular frequency $\omega_{\rm peri}$ at
 pericentre, and we restrict ourselves in this work to unbound orbits
 only. We choose $\omega_{\rm peri}$ such that a prescribed number of
 inner Lindblad resonances is present in the disc, i.e. that certain
 orbits in the disc are in resonance with the companion, when passing
 $R_{\rm peri}.$\footnote{We point out that the angular frequency of
 the companion, and therefore also the location of the corresponding
 resonances in the disc, changes with time.} Since the stellar bar in
 the unperturbed host galaxy is very weak, we apply the linear
 approximation, as described in Sec.~\ref{hostI0}, in order to
 determine the resonances in the disc. The resonance diagram for the
 isolated model I1 is shown in Fig.~\ref{fig12}, and the frequencies
 chosen for the companion are:
\begin{tabbing}
 1. \quad \=  $\omega_{\rm peri}\!=\!0.3$ \quad \= (two ILRs) \\
 2.       \>  $\omega_{\rm peri}\!=\!0.6$       \> (one ILR)  \\
 3.       \>  $\omega_{\rm peri}\!=\!0.9$       \> (no ILR) \ . 
\end{tabbing}
 The mass of the companion for these simulations has been varied, as
 before, from $1\!:\!1$ to $1\!:\!4$, and we have a few additional
 runs with even lower mass ratios, namely $1\!:\!\frac{1}{2}$ (C6)
 and $1\!:\!\frac{1}{3}$ (C5). A complete list of all simulations
 and of their main parameters is given in Tab.~\ref{tab03}. Some of
 the companion orbits are illustrated in Fig.~\ref{fig13} for
 different eccentricities and pericentric separations.

\begin{figure*}
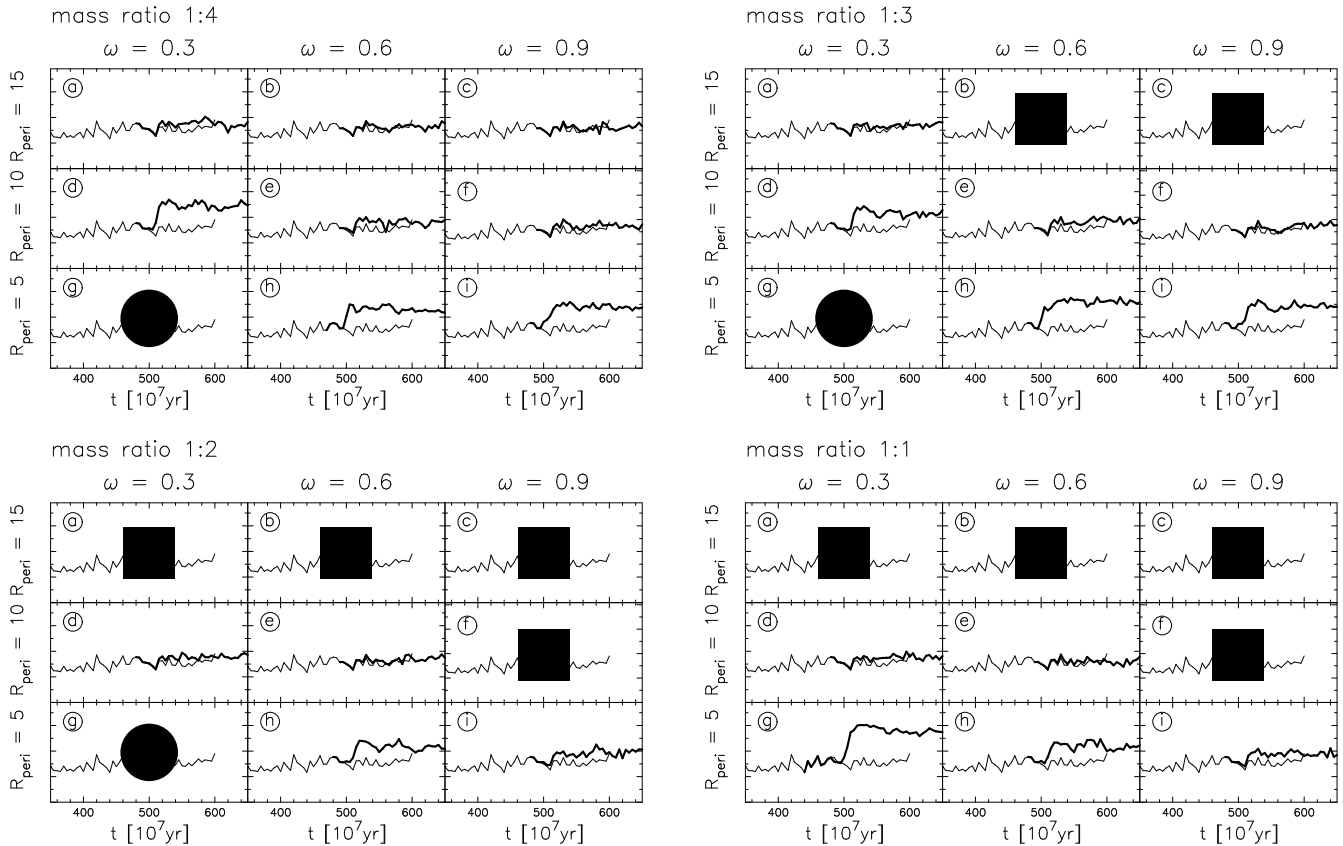
 
 \includegraphics[angle=-90, width=\columnwidth]{fig14a} \hspace{5ex}
 \vspace{3ex}
 \includegraphics[angle=-90, width=\columnwidth]{fig14b}
 \vspace{3ex}
 \includegraphics[angle=-90, width=\columnwidth]{fig14c}
 \hspace{5ex}
 \includegraphics[angle=-90, width=\columnwidth]{fig14d}
 \caption{Bar strength as a function of time for some of the pure
     stellar models.  The four frames show the models with mass
     ratios of $1\!:\!1$, $1\!:\!2$, $1\!:\!3$ and $1\!:\!4$,
     respectively. The filled squares mark models which have not been
     run, because other runs allow us to deduce that a bar cannot
     form. Models in which the companion would be on a bound
     elliptical orbit are marked by a filled circle.}
 \label{fig14}
\end{figure*}

 In Fig.~\ref{fig14} we show the bar strength as a function of time
 for this set of interaction models. It is striking that -- in
 contrast to the models with gas -- the interaction is sufficiently
 strong to regenerate the bar in about half our models.  As can also
 be seen from the plot, there is a general trend, that the strength
 of the induced bar increases with
 \begin{itemize}
  \item increasing mass $M_{\rm C}$ of the companion,
  \item decreasing pericentric separation $R_{\rm peri}$,
  \item decreasing pericentric frequency $\omega_{\rm peri}$.
 \end{itemize}
 This is what one might generally expect, since the gravitational
 force of the companion is proportional to its mass, the tidal
 force decreases with  $R^{-3}$, and the time integrated force of
 the companion on the host galaxy becomes stronger for slower
 passages.
 
 Furthermore, after the bar has formed in the disc, we find that the
 amplitude of the bar stays constant with time. We have checked the
 life-time of the regenerated bar by running model I1\,C4\,d, which
 shows a significant increase in bar strength after the interaction,
 for roughly $\Delta t\!=\!4\times10^{10}$\,yr in total. Using a
 linear-fitting we find a decay-rate of the bar of roughly
 $9\!\times\!10^{-12}$\,yr. With this the bar will drop to half its
 amplitude after approximately $\Delta t\!=\!5\!\times\!10^{10}$\,yr.
 After one Hubble time (with $H_0^{-1}\!=\!1.3\!\times\!10^{10}$\,yr)
 the bar amplitude would have dropped only by 12 per cent. A fraction
 of this decrease could be introduced by the relatively low
 number of particles used in the simulations, so that in fact there
 could be (almost) no sign of a decay for bars formed in
 interactions. These results strongly argue for the fact that bars
 formed by the interaction are long-living and by no means
 transient phenomena.

\subsection{Dynamical properties of the bars}

\begin{figure} 
\begin{center}
 \includegraphics[angle=-90, width=\columnwidth]{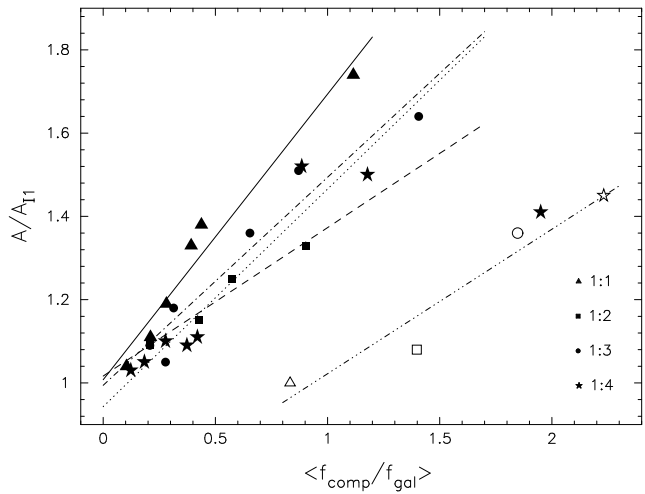}
\end{center}
 \caption{Normalized bar strength A/A$_{\rm I1}$ versus the interaction
   strength parameter $\Theta$. We show the results of the purely
   stellar simulations and the corresponding linear fits for mass
   ratios of $1\!:\!1$ (bullets, full line), $1\!:\!2$ (squares,
   dashed line), $1\!:\!3$ (triangles, dotted-dashed line) and
   $1\!:\!4$ (stars, dotted line). The simulations with parabolic
   orbits are shown with open symbols and are not taken into account
   for the linear fits.}
 \label{fig15}
\end{figure}

 To quantify the correlations described in the previous section, we
 define a parameter $\Theta$, which allows us to evaluate the
 interaction strength:
\begin{equation}
 \Theta \equiv
 \left< \frac{ f_{\rm comp}}{ f_{\rm gal} } \right>  =
   \int \frac{ f_{\rm comp} }{ f_{\rm gal} }\  {\rm d}t ,
 \label{eq04}
\end{equation}
 where $f_{\rm gal}$ and $f_{\rm comp}$ are the mean radial forces,
 exerted from the host I1 and the companion galaxy, respectively,
 averaged over the equatorial plane of the host galaxy. In practice
 we calculate them on a radial equally-spaced polar grid with maximum
 radius $R\!=\!4.8$ and then take a density weighted average. The
 integral in eq. (\ref{eq04}) is carried out from $t\!=\!-250$ to
 $t\!=\!250$, with $t\!=\!0$ corresponding to $t_{\rm peri}$, 
 in order to guarantee an adequate convergence of $\Theta$ in all
 models. For these force calculations both galaxies are approximated
 by point masses for simplicity. The results for $\Theta$ are given
 in the penultimate column of Tab.~\ref{tab03}. We also calculate the
 normalized bar strength $A/A_{\rm I1}$, where $A$ and $A_{\rm I1}$
 are the bar strength in the interaction model and in the isolated
 model, respectively. By plotting the normalized bar strength versus
 the interaction strength parameter $\Theta$ (Fig.~\ref{fig15}), we
 find a roughly linear correlation between the two quantities for
 each companion C$i$, with $i\!=\!1\ldots 4$. One of the $1\!:\!4$
 models (I1\,C4\,h) is systematically off-set in all plots of this
 kind (i.e., Figures~\ref{fig15}, \ref{fig17} and \ref{fig19}), but
 always lies within the $2\sigma$ confidence limit. The models with
 parabolic encounters are all off-set from the relationships found,
 but seem to be correlated themselves linearly.

\begin{figure} 
 \begin{center}
 \includegraphics[angle=-90, width=\columnwidth]{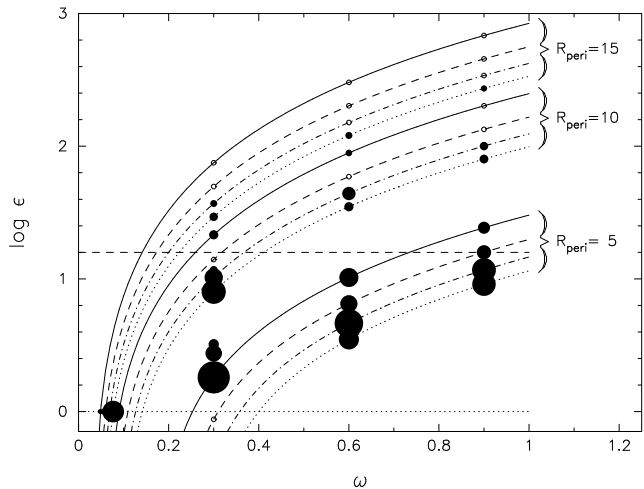}
 \end{center}
 \caption{Logarithmic orbital eccentricity versus the
   angular frequency of the companion at pericentre time. For each 
   pericentric separation the curves obtained from the corresponding
   two-body problem are plotted for different mass ratios:
   $1\!:\!1$ (full line), $1\!:\!2$ (dashed), $1\!:\!3$
   (dotted-dashed) and $1\!:\!4$ (dotted line). The filled circles
   mark the different simulations and their size is proportional to
   the bar strength at the end of each run. The two circles which do
   not lie on a line are simulations with mass ratios of $1\!:\!0.5$
   and $1\!:\!0.33$. The limit of $\epsilon$, as explained in the text,
   is marked by the horizontal dashed line.}
 \label{fig16}
\end{figure}

\begin{figure} 
\begin{center}
 \includegraphics[angle=-90, width=\columnwidth]{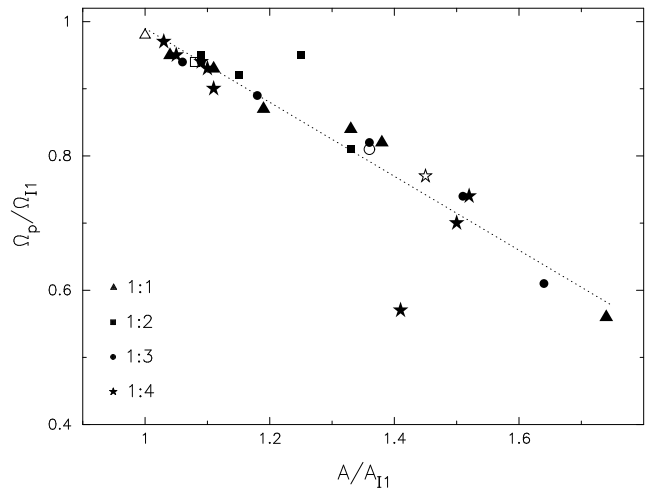}
\end{center}
 \caption{Normalized pattern speed $\Omega_{\rm p}/\Omega_{\rm I1}$
   versus normalized bar strength $A/A_{\rm I1}$ of the purely
   stellar interaction models. Symbols are as in Fig.~\ref{fig15}.}
 \label{fig17}
\end{figure}

 Plotting the bar strength of the different models in a diagram
 showing the logarithmic orbital eccentricity $e$ versus the angular
 frequency $\omega_{\rm peri}$ (see Fig.~\ref{fig16}), we find that
 models with orbits of roughly $\log e\!>\!1.2$ are not sufficiently
 strong to regenerate the bar in the purely stellar models. A larger
 sample of simulations has to be run, however, to confirm the
 existence of such a limiting eccentricity. If such a limit really
 exists, it would give a necessary, but not sufficient condition for
 tidally induced bar formation for our purely stellar model.

 Angular momentum exchange in isolated galaxies leads to correlations
 between the bar pattern speed and strength, as well as between the
 bar strength and the angular momentum gained by the spheroid, which,
 in many cases, is a measure of the angular momentum exchanged \cite{ath03}. We
 will now test whether such correlations can be found in our
 regenerated bars. We indeed find a tight correlation between the
 strength and the pattern speed of the regenerated bars
 (Fig.~\ref{fig17}). The stronger the bar gets after the interaction,
 the lower becomes its pattern speed, in good agreement with
 Athanassoula \shortcite{ath03}. We further find that the pattern
 speed in the interaction models is always lower than in the
 corresponding isolated case.

\begin{figure} 
 \begin{center}
 \includegraphics[angle=-90, width=\columnwidth]{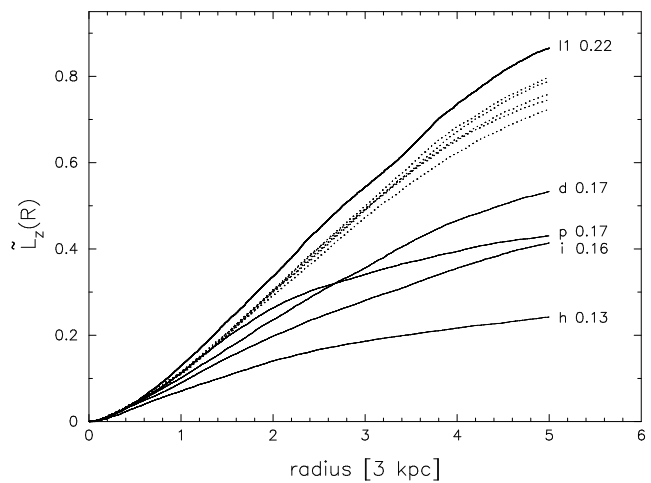}
 \end{center}
 \caption{Specific cumulative angular momentum of the stellar disc
   for the isolated model I1 (thick line) and different interaction
   models I1\,C4 (thin lines). The dotted lines indicate models in
   which no or only a weak increase of bar strength has been found
   after the interaction. Models in which the bar has been
   regenerated (full lines) are labelled (with the orbit type as
   given in Tab.~\ref{tab03}) and the corresponding bar pattern speed
   is given.}
 \label{fig18}
\end{figure}

\begin{figure} 
 \begin{center}
 \includegraphics[angle=-90, width=\columnwidth]{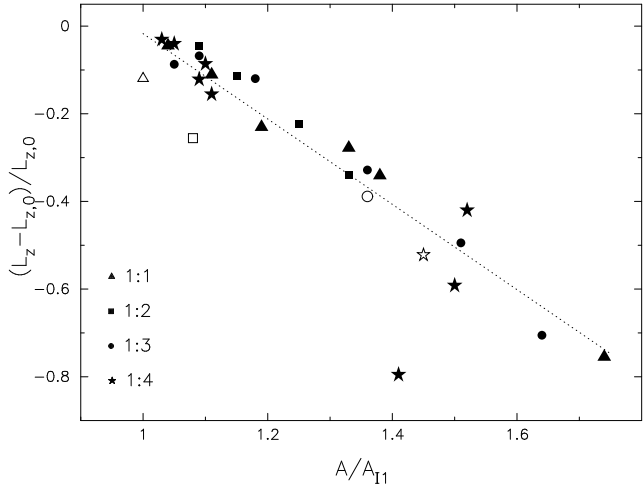}
 \end{center}
 \caption{Total change of angular momentum $\Delta L_z$ of the
   stellar disc inside its initial cut-off radius as a function of
   bar strength normalized by the corresponding value of the isolated
   model. The different symbols represent interactions with different
   mass ratios. Parabolic and hyperbolic orbits are represented by 
   open and filled symbols, respectively.}
 \label{fig19}
\end{figure}

 The passage of the companion again is accompanied by a
 redistribution of disc angular momentum. The angular momentum
 exchange between the different disc regions (as defined in the 
 previous section) is similar to that in Fig.~\ref{fig08}. In
 Fig.~\ref{fig18} we show the radial distribution of the specific
 cumulative angular momentum ${\tilde L}_z(R)$ for a set of
 simulations with companion C4 in comparison with the isolated model
 I1. For the runs with lower mass companions, we find similar
 results. It is noticeable that the regeneration of the stellar bar
 has been feasible only in those models which show a significant
 change in ${\tilde L}_z$. Actually we find a tight correlation
 between the loss of angular momentum $\Delta L_z$ in the disc
 measured inside the initial disc cut-off radius $r_{\rm cut}$ and
 the induced bar strength, as shown in Fig.~\ref{fig19}. The more
 angular momentum is removed from the disc, the stronger the
 regenerated bar becomes, in good agreement with what was found for
 isolated bars by Athanassoula \shortcite{ath03}. We find the same
 correlation about equally strong when we plot the angular momentum
 change within the corotation radius. This important connection
 between the angular momentum exchange and the (re)generation process
 of the bar is described further in the discussion section.

\section{Discussion} \label{discussion}

\begin{figure}
\begin{center}
 \includegraphics[angle=-90,width=\columnwidth]{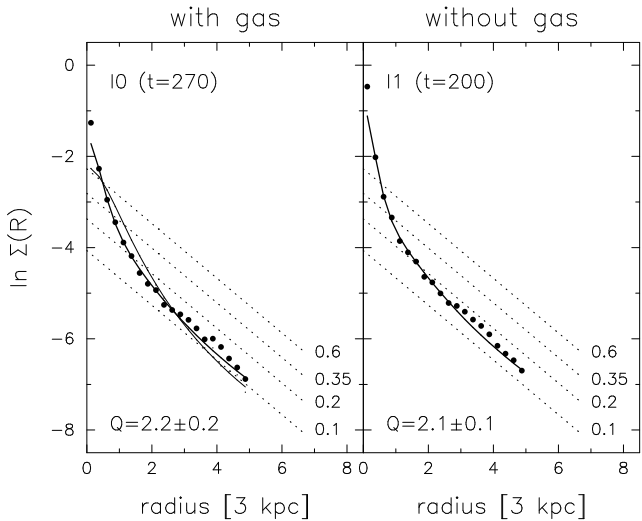}
\end{center}
 \caption{Logarithmic surface density of the stellar disc of models
   I0 (left-hand panel) and I1 (right-hand panel).  The filled dots
   show the density distribution of the stellar discs and the thick
   full lines the best-fit of the corresponding Kuzmin-Toomre
   profile.  The thin full line in the left-hand panel gives the
   density distribution of the initial stellar disc in model
   I0. We also show the different exponential discs used by Miwa
   \& Noguchi \shortcite{mn98} with dotted lines and give the mass of
   each disc in their model units. The value of the Toomre
   Q-parameter of the stellar disc is given at the bottom of each
   panel.}
 \label{fig20}
\end{figure}

 The regeneration of stellar bars triggered by galaxy interactions
 has been suggested as an additional scenario for the formation
 of bars and for explaining the observed frequency of bars along the
 Hubble sequence (Sellwood \& Moore  1999 ; Friedli 1999;
 Athanassoula 2000). So far only the formation of tidally induced
 bars in initially non-barred disc galaxies and the required
 conditions, under which such an event may occur, have widely been
 studied by means of numerical simulations (Byrd et al. 1986; Noguchi
 1988; Salo 1991; Miwa \& Noguchi 1998; etc.). These 
 results, however, need not necessarily apply to interactions with a
 {\em formerly} barred galaxy. Indeed, a former stellar bar
 likely might have changed the dynamics of the disc, i.e. increased
 the velocity dispersion in the stellar disc and changed both the
 density and angular momentum distribution (compare Fig.~\ref{fig20}).
 Therefore, the regeneration of stellar bars is subject to different
 conditions than the formation of a bar in an bar-unstable isolated
 disc or in a tidal interaction.

 In this work we present numerical simulations starting with a host
 galaxy which has initially been bar unstable and in which the bar
 has been significantly weakened due to gas inflow before the
 interaction. 

\subsection{Quantifying the interaction strength} \label{barprop}

 To quantify the strength of the interaction, we have introduced a
 parameter $\Theta$ for the interaction strength (Sec. 6.1), which
 basically takes the following quantities into account: 1) the mass
 of the companion, 2) the pericentric separation between the
 galaxies, and 3) the velocity of the companion at pericentre,
 defining a kind of interaction time-scale. With our definition (see
 Sec.~6.1), we find that the interaction strength $\Theta$ correlates
 well, i.e. roughly linearly, with the strength of the regenerated
 bars, which have been created in our purely stellar models.

\begin{figure}
\begin{center}
 \includegraphics[angle=-90, width=\columnwidth]{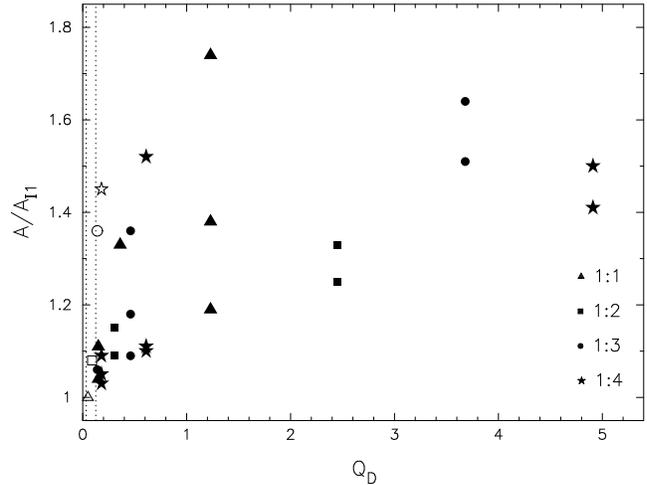}
\end{center}
 \caption{Normalized bar strength A/A$_{\rm I1}$ versus the Dahari
    index $Q_{\rm D}$ (Dahari 1984) for interactions with different
    mass ratios. The layout is as in Fig.~\ref{fig15}.}
 \label{fig21}
\end{figure}

 A different parameter for the interaction strength has been
 introduced by Dahari \shortcite{dah84} as an estimate of the direct
 tidal impulse:
\begin{equation}
 Q_{\rm D}\!=\!(M_{\rm comp}/M_{\rm host}) / 
               (R_{\rm min}/R_{\rm disc})^3,
\end{equation}
 where $M_{\rm comp}$ and $R_{\rm peri}$ are the mass of the
 companion galaxy and the pericentric separation, respectively.
 $M_{\rm host}$ is the mass of the host galaxy within the disc
 truncation radius $R_{\rm disc}$. Salo \shortcite{sal91} has used
 the Dahari index as a quantitative measure for tidally induced bar
 formation in his 2D simulations and found a minimum value of
 $Q_{\rm D}$, depending on the specific host galaxy, above which a
 bar is formed by the interaction. The specific limiting value of
 $Q_{\rm D}$ Salo found depends strongly the central mass
 concentration of the host galaxy and cannot be easily transferred to
 our model. Furthermore, applying the Dahari index to our simulations
 we find neither a correlation between the strength of the
 interaction and the regeneration of the bar, nor a limiting value of
 Q$_{\rm D}$ (see Fig.~\ref{fig21}). The main drawback of the Dahari
 index is that it does not take into account the interaction orbit,
 i.e. the interaction time-scale. The Dahari index therefore does not
 seem to be sufficient to constrain the parameters necessary for the
 formation or regeneration of bars by galaxy interactions in general.

\begin{figure} 
\begin{center}
 \includegraphics[angle=-90, width=\columnwidth]{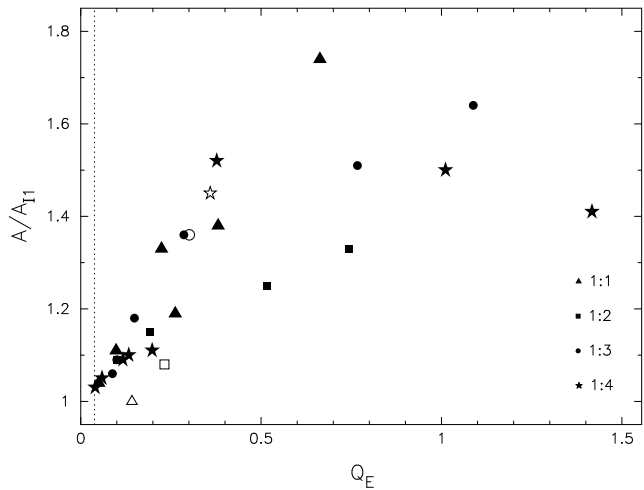}
\end{center}
 \caption{Normalized bar strength versus the interaction parameter
    $Q_{\rm E}$ (Elmegreen et al. 1991). The layout is as in
    Fig.~\ref{fig15}.}
 \label{fig22}
\end{figure}

 A more advanced parameter has been introduced by Elmegreen et al.
 \shortcite{elm91}. This is based on the Dahari index, but includes
 also the ratio between an interaction time-scale $\Delta T$ and some
 dynamical time $T$ of the disc:
\begin{equation}
   Q_{\rm E}=Q_{\rm D} \times \frac{\Delta T}{T} ,
\end{equation}
 where $T\!=\!R^3_{\rm gal}/{\rm G} M_{\rm gal}$ and $\Delta T$ is
 the time it takes the companion to move by one radian relative to
 the hosts centre at pericentre time. These authors also report a
 limiting value $Q_{\rm E}\!=\!0.038$ for the formation of the bar by
 tidal interactions. Applying this parameter to our simulations we
 find a correlation between the interaction strength and the strength
 of the regenerated bar (see Fig.~\ref{fig22}). This, however, is
 less pronounced than the corresponding one for our parameter
 $\Theta$, which takes more fully the interaction into account. The
 existence of a limiting parameter, as found by Salo
 \shortcite{sal91} and Elmegreen et al. \shortcite{elm91}, however,
 might be attributed to the method for measuring the bar strength.
 For our purely stellar simulations we find a correlation between
 $\Theta$ and the induced bar strength rather than a limiting value
 of $\Theta$. Still, the exact correlation between $\Theta$ and the
 bar strength -- as described in Sec.~\ref{interI1}.1 -- is likely to
 be model dependent.

\subsection{Radial redistribution of angular momentum}

 Athanassoula \shortcite{ath03} argued that the redistribution of
 angular momentum is the driving agent for the evolution of an
 isolated barred galaxy. Both her analytical work and her numerical
 simulations show that galaxies that have exchanged more angular
 momentum should have a stronger bar, with a faster decreasing
 pattern speed. In our purely stellar models, we also find that,
 whenever bar regeneration occurs, it is always accompanied by a
 considerable loss of angular momentum from the disc (see
 Fig.~\ref{fig18}).
 Indeed, as in the case of isolated galaxies \cite{ath03}, there is a
 tight correlation between the strength of the bar and the angular
 momentum lost by the inner disc. This argues that angular momentum
 exchange is tightly linked to the bar formation process, independent
 of whether that is spontaneous, or driven.

 The actual exchange process, however, is not always the same. In
 isolated discs angular momentum is emitted by particles in resonance
 with the bar in the inner part of the disc (within corotation), and,
 to a lesser extent and if there is a considerable bar growth, by
 non-resonant particles in that region. This is absorbed by particles
 in the outer disc and halo. In this case there is only one pattern
 speed, that of the bar, since the spirals, which could in principle
 have a different pattern speed (Tagger et al. 1987; Sygnet et al
 1988), have died away in the early parts of the simulations. Thus
 resonances are well defined. This is certainly not the case here.
 Besides the bar pattern speed, there is the driving frequency
 of the interaction which changes with time.
 Furthermore, the forcing from the companion is of comparable
 strength to that of the bar, and the companion itself participates
 actively in the angular momentum exchange. Thus, in contrast to the
 isolated cases, we do not find the bar corotation radius to separate
 disc angular momentum emitters from disc absorbers. 

\begin{figure*}
\begin{center}
 \includegraphics[angle=-90, width=\textwidth]{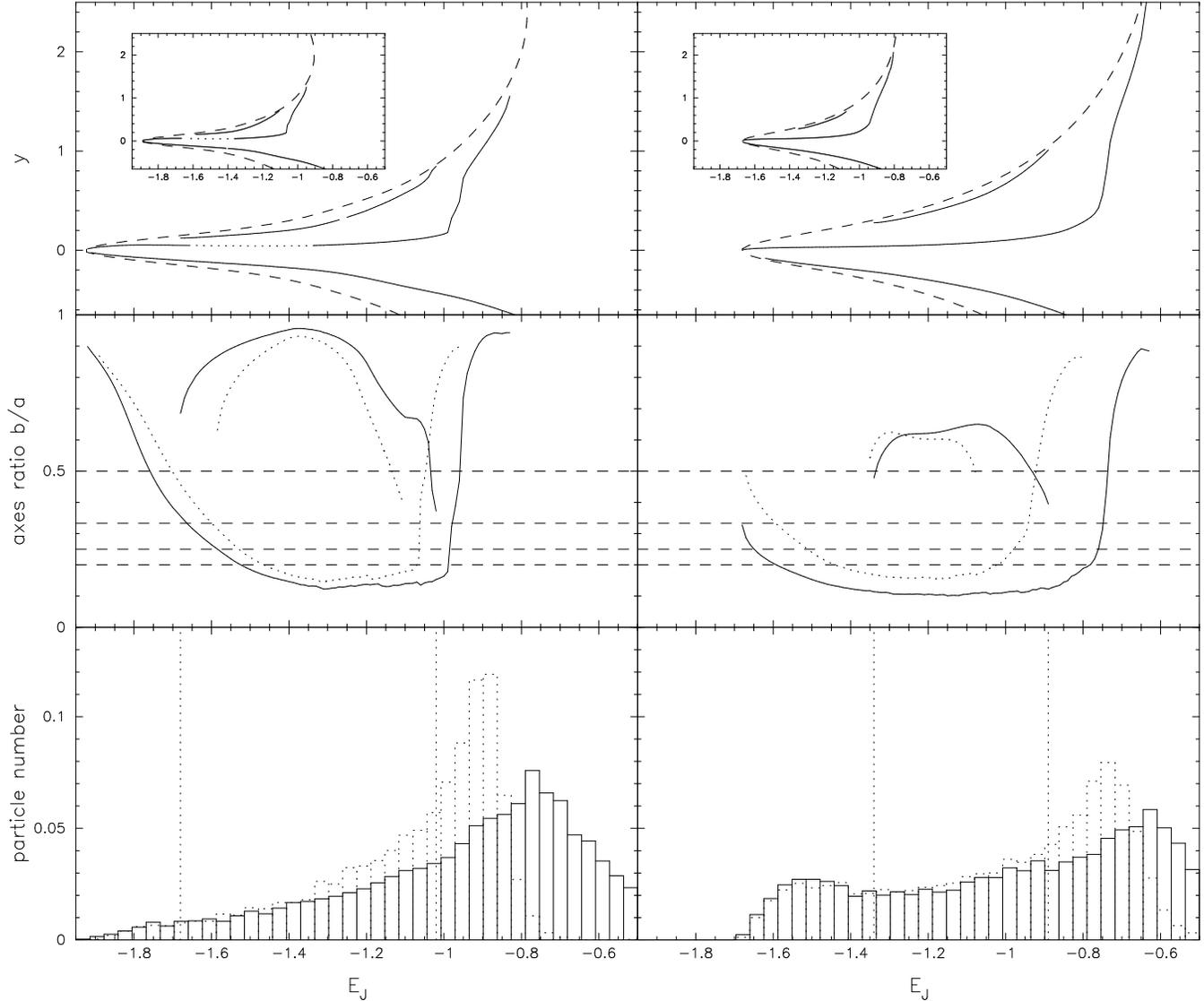}
\end{center}
 \caption{Comparison of models I0\,C4\,p (with gas; left-hand side)
  and I1\,C4\,p (without gas; right-hand side). The upper panels
  show the characteristic diagram at the end of the run of the
  interaction model (main panel) and of the corresponding isolated
  model (subpanel). The layout of the diagrams is the same as for,
  e.g., Fig.~\ref{fig03}. Instability regions in the characteristic
  curves are indicated by dotted lines.
  In the middle panels we show the axial ratios $b/a$ of the $x_1$
  and $x_2$ orbit families, where $a$ and $b$ are the major and minor
  axis, respectively. The full and dotted lines show the results of
  the interaction and the isolated model, respectively. The
  horizontal dashed lines indicate axes ratios of $1\!:\!2$,
  $1\!:\!3$, $1\!:\!4$ and $1\!:\!5$.
  The bottom panels show a histogram of the number of stellar disc
  particles within corotation per energy interval, normalized by
  total number of stellar disc particles within corotation in the
  corresponding isolated model. The vertical dotted lines mark the
  region in which the $x_2$ orbits is present in the perturbed
  models.}
 \label{fig23}
\end{figure*}

 The bar regeneration process can be be understood in terms of
 basically the same angular momentum considerations as described
 by Athanassoula \shortcite{ath03} for isolated disc galaxies.
 The bar in an isolated disc can get stronger and lose angular
 momentum basically by four different effects, which should be
 linked to each other. First, particles which were on quasi-circular
 orbits outside the bar get trapped into elongated orbits in its
 outer part and thus the bar becomes longer. Secondly, the orbits
 trapped in the bar could get thinner and make the bar 
 thinner, too. Third, more mass could be get trapped on
 periodic orbits in the bar. And finally, the bar can of course lose
 angular momentum by slowing down. These effects should also be present
 in our case as well, where the loss of disc angular momentum is
 predominantly driven by the tidal perturbation of the companion. We
 will discuss here how much the different effects contribute to the
 regeneration of the stellar bars in our simulations. Since 
 analytical calculations cannot make any statements about this, we
 will base the discussion on the dynamical properties and orbital
 structure of our specific models.

\begin{figure}
\begin{center}
 \includegraphics[angle=-90, width=\columnwidth]{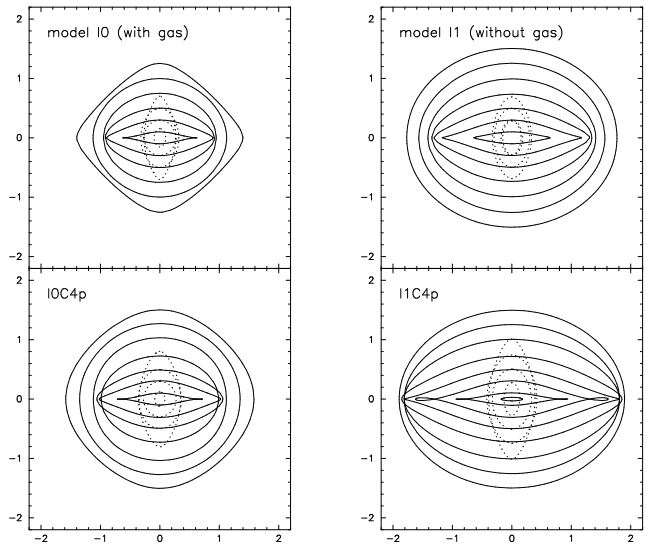}
\end{center}
 \caption{Concentric orbits of the $x_1$ (full lines) and $x_2$
   (dotted lines) families.  The left- and the right-hand column
   show the models with and without gas, respectively. In the upper
   panels we show the isolated models I0 and I1 and in the lower
   panels models I0\,C4\,p and I1\,C4\,p.}
 \label{fig24}
\end{figure}

 In Fig.~\ref{fig23} (right-hand column, upper panel) we show a
 direct comparison of the characteristic diagrams of the purely
 stellar models I1 and I1\,C4\,p at the end of each run, respectively.
 The characteristic diagram of the isolated and the perturbed case
 look very similar in terms of the layout of the main periodic
 orbits. The main difference, however, is that both the $x_1$ and the
 $x_2$ orbits extend to higher energies after the interaction. This
 is especially obvious for the part of the $x_1$ characteristic
 around $E_{\rm J}\!=\!-0.75$, where the value of the $y$ intercept
 increases strongly with $E_{\rm J}$. This clearly moved towards
 higher energies. In Fig.~\ref{fig24} (right-hand column) we show
 some examples of the periodic orbits with different Jacobian
 energies $E_{\rm J}$ of both the $x_1$ and the $x_2$ family. We
 find that the $x_1$ orbits become longer and, to some smaller extent
 though, also thinner after the interaction. This becomes clearer in
 Fig.~\ref{fig23} (middle panel), in which we plot the axial ratio
 $b/a$ of the orbits, where $a$ and $b$ denote the major and minor
 axis, respectively, as a function of $E_{\rm J}$. The main effect,
 however, is the lengthening of both the $x_1$ and the $x_2$ orbits
 towards higher energies and, taking into account the corresponding
 characteristic diagram, also in radial extent. For the $x_1$ orbits
 we also notice some lengthening towards lower energies, as well.
 Owing to these effects the bar gets both more centrally concentrated
 and more extended to large radii.

 The thinning of the periodic orbits, which is less pronounced
 in our models than the lengthening, can be understood by the
 analytic calculations by Lynden-Bell (1979; hereafter LB79) in the
 context of gradual bar growth in isolated disc galaxies. As
 described in LB79, the mean circular frequency $\Omega_{\rm c}$ of
 most disc stars in the central and/or inner disc region is much
 higher than the pattern speed $\Omega_{\rm p}$ of the weak periodic
 perturbation, which is considered to be bar-like in this case. In
 the frame of reference corotating with the perturbation, the fast
 motion $\Omega_{\rm c} - \Omega_{\rm p}$ of the stars on
 near-circular orbits is not considerably affected by the weak
 perturbation and therefore the fast action variable $J_{\rm f}$
 (see LB79) is approximately constant. As shown by Lynden-Bell,
 angular momentum is removed from the orbits, whose elongation leads
 the perturbing bar-like potential. Near-resonant orbits in the inner
 disc region, which reside in the abnormal region (see Fig.~2 in
 LB79), i.e. the region where
 $\Omega_i\!=\!\Omega_{\rm c}-\frac{1}{2}\kappa$ of the orbit
 decreases as its angular momentum decreases with constant
 $J_{\rm f}$, will align with the perturbation and become more
 eccentric, supporting the growth of the bar structure. As pointed
 out by Lynden-Bell, this way the bar shape also becomes more
 eccentric, but the length of the bar will not change significantly.
 The main difference to the scenario described in LB79 is that in our
 case, the perturbation is not periodic, since it is due to both the
 companion and the bar, and the angular frequency of the former is a
 function of time. 

 We finally check which orbits are important for making the bar. In
 the lower panel of Fig.~\ref{fig23} we therefore plot the number of
 particles as a function of $E_{\rm J}$, normalized by the total disc
 particle number within corotation of the isolated model I1. This
 way we get information about how the principal orbits at different
 energies are populated in our model. As can be seen from this plot,
 the total number of stellar particles within the corotation radius
 increased by less than 10 per cent. We conclude that
 the mass of the bar does not increase significantly. The main effect
 we find is that particles with energies close to the vertical
 $x_1$-branch move towards higher energies or higher radial extent,
 taking into account the characteristic diagram.

\begin{figure*}
\begin{center}
 \includegraphics[angle=-90, width=.8\textwidth]{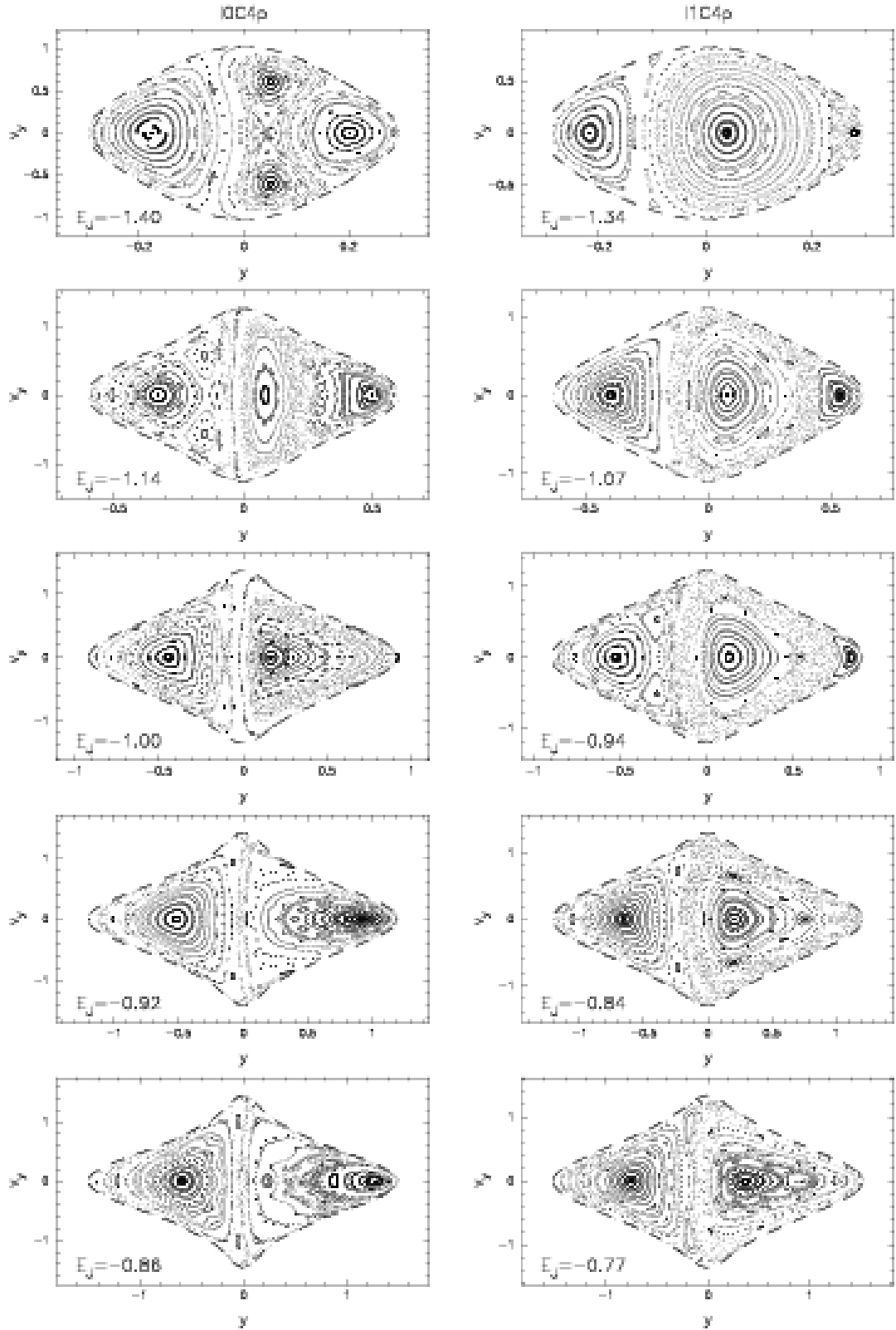}
\end{center}
 \caption{Surfaces of section for models I0\,C4\,p (with gas;
 left-hand column) and I1\,C4\,p (without gas; right-hand column) at
 the end of the runs. The values of the Jacobian energy $E_{\rm J}$,
 given in the bottom left corner of each panel, have been chosen such
 as to give roughly the same range in $y$ for each row. In the
 purely stellar model the phase space is dominated by the $x_1$ and
 $x_4$ orbits at low Jacobian energies,
 while in the models with gas the phase space is mainly
 dominated by the $x_2$ and irregular orbits.}
 \label{fig25}
\end{figure*}

 By tracking individual particle orbits in the corresponding
 potential and plotting their average y-intercept versus the Jacobian
 energy $E_{\rm J}$ in the characteristic diagram, we find that the
 $x_2$ orbits in the purely stellar models are only populated by a
 very small fraction of particles. This result is consistent with
 Fig.~\ref{fig25}, in which we show the surfaces of section of
 model I1\,C4\,p at the end of the run for different $E_{\rm J}$. As
 can be clearly seen, the $x_1$ family dominates phase space over a
 wide range in energy and the $x_2$ orbits are almost not present.
 
\subsection{Influence of the gas}

 Interactions which are strong enough to induce a bar in our purely
 stellar models are not sufficiently strong to induce a bar in the
 $N$-body/SPH models. In fact it has not been possible to induce a
 bar in any of our dissipative models, and we will here discuss the
 possible reasons for this. The stellar disc component in these
 simulations loses angular momentum, as was found also in the purely
 $N$-body models (see Fig.~\ref{fig08} and \ref{fig18}). The gas 
 in the inner parts, however, gains angular momentum, so that  
 the inner parts of the disc may gain a small amount of angular
 momentum (see Fig.~\ref{fig09}), or if they lose, it is considerably
 less than the corresponding quantity for the purely stellar case.
 Furthermore, as was shown in Fig.~\ref{fig07} the interaction is
 accompanied by a significant inflow of gas towards the central disc.
 These two effects, coupled, prevent the regeneration of the bar. In
 this section we discuss how the increase of the central mass
 concentration affects the orbital structure of the disc and thus
 helps prevent the regeneration of the bar in these models.

 In Fig.~\ref{fig23} (left-hand column) we compare models with gas,
 before and after the interaction. In particular, for models I0
 and I0\,C4\,p, we compare the characteristic diagrams (upper panel),
 the axial ratio of the $x_1$ and $x_2$  orbits (middle panel) and
 the particle number per energy range (bottom panel). In contrast to
 the models without gas (right column) the characteristic diagram of
 the models with gas extends more to lower energies, as would be
 expected from Fig.~\ref{fig07} and Sec.~\ref{hostI1}, that show
 clearly that the density distribution is centrally more concentrated
 in models with gas. Furthermore, the extent of the $x_2$ orbits is
 considerably larger, and the $x_1$ orbits have a sizeable
 instability strip between roughly $E_{\rm J}\!=\!-1.34$ and $-1.66$. 

 Information on the axial ratio of the $x_1$ and $x_2$ orbits and on
 their extent is given in the middle panels of Fig.~\ref{fig23} and 
 in Fig.~\ref{fig24}. We see clearly that the differences between the
 cases before and after the interaction are much smaller than for the
 purely stellar case. The $x_1$ orbits are a bit thinner after the
 interaction and the $x_2$ orbits have slightly larger radial extent,
 but the differences are small. From Fig.~\ref{fig23} and \ref{fig24}
 we see that the $x_1$ orbits are less elongated than the
 corresponding orbits in the models without gas. This is particularly
 true at the highest and lowest energies. The extent of the elongated
 orbits is also considerably shorter. Also the $x_2$ orbits in the
 central region, i.e. at lower energies, are much rounder in the case
 with gas. 

 The number of stellar particles within corotation in models with
 gas (Fig.~\ref{fig23}; bottom panel) have increased by roughly 12
 per cent compared to the corresponding isolated model. Particles
 move from the intermediate energy region to the high energy one, and
 also, though to a lesser amount, to the low energy central region,
 similar to what is found in the models without gas. 

 Comparing the SOSs of models with and without gas (I0\,C4\,p and
 I1\,C4\,p, respectively) after the interaction (Fig.~\ref{fig25}), we
 note that the area corresponding to $x_1$ orbits is much smaller in
 models with gas. On the other hand the area corresponding to chaotic
 motion and the area corresponding to $x_2$ orbits is considerably
 larger. The chaos in the cases with gas is due to the instabilities
 of the $x_1$ orbits, discussed above. The larger area covered by the
 $x_2$ orbits is in agreement with the fact that the $x_2$
 characteristic is much more extended. To pursue this further we
 followed in the frozen potential the orbits of particles with
 initial conditions from the simulation, as we had already done for
 the simulations without gas. We find a notable difference. Namely
 there is now an indication that, contrary to the purely stellar case, a
 non-negligible fraction of the orbits is trapped around $x_2$ orbits.
 This difference can be understood as a result of the induced
 gas-inflow (see Fig.~\ref{fig07}), in accordance with the results
 found in numerical simulations of isolated gas-rich barred galaxies
 (e.g., Friedli \& Benz 1993; Berentzen et al. 1998). Owing to
 the growing central mass concentration the $x_2$ orbits cover a
 larger phase-space volume at the expense of the $x_1$ orbits.

 Thus the fact that bars cannot be regenerated in simulations with gas
 can be understood with the help of the many differences between the
 two models, described above. There is considerably less angular
 momentum loss from the inner disc material, if this is not a gain. 
 The interaction brings considerably less change of the shape and
 extent of the $x_1$ orbits, but renders them unstable over a
 considerable energy interval, thus introducing a considerable
 amount of chaos. Finally, due to the increased central concentration,
 the importance of the $x_2$ population is considerably increased. The
 coupling of these three anti-bar effects prevents the regeneration of
 a bar component.
 
\subsection{Properties of regenerated bars}

 As a next step we discuss the properties of the regenerated bars
 compared to the ones formed by the bar instability in isolated
 discs. Contrary to Miwa \& Noguchi \shortcite{mn98}, we find no clear
 evidence for qualitative differences between the two types of bars,
 while we do find clear evidence for similarities. The orbital
 differences between the two types of bars, discussed in the two
 previous subsections are quantitative, rather than qualitative, since
 they pertain only to the extent of the families and the shape of the
 corresponding orbits. The role of the resonances seems the same.

\begin{figure}
\begin{center}
 \includegraphics[angle=-90, width=\columnwidth]{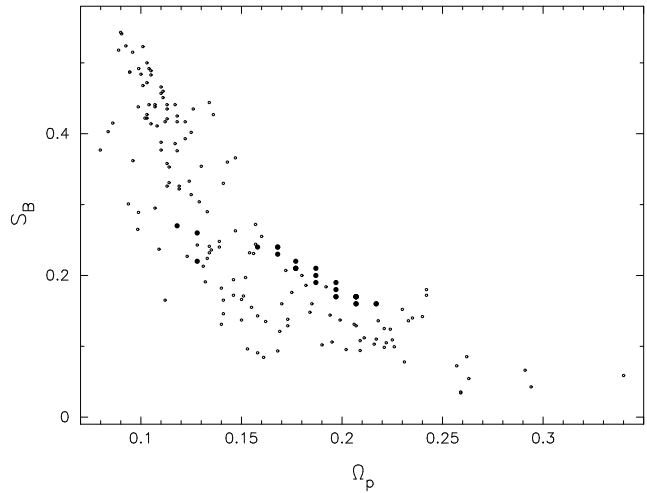}
\end{center}
 \caption{Bar strength $S_{\rm B}$ versus pattern speed $\Omega_{\rm p}$
  in units used by Athanassoula \& Misiriotis (2002). The dots and
  bullets show the results of the isolated models of Athanassoula (2003)
  and our purely stellar interaction models, respectively.}
 \label{fig26}
\end{figure}

 Athanassoula \shortcite{ath03} found a correlation between the bar
 strength and the pattern speed of bars in isolated disc galaxies.
 In order to compare the properties of the bars formed in our models
 and the ones in isolated discs, we first calibrate our model units
 appropriately, as proposed in Athanassoula \& Misiriotis
 \shortcite{am02}, and then compared them to the ones in Athanassoula
 \shortcite{ath03}. The result is shown in 
 Fig.~\ref{fig26}, which shows that the area covered by the driven
 bars is not separated by that covered by the isolated bars. This
 argues strongly for the similarity between the dynamical properties
 of the two types of bars, and the difficulty to distinguish between
 them.

 Elmegreen \& Elmegreen \shortcite{ee85} found two different types of
 major axis surface density profiles, the exponential and the flat
 ones, and Noguchi \shortcite{nog96} argued that this distinction
 could be due to the fact that the first type is found in spontaneous
 bars and the second in driven ones. We believe that the difference
 is due to the bar strength, rather than its origin, since
 Athanassoula \& Misiriotis \shortcite{am02} showed that both
 types of profiles could be found in isolated galaxies: the flat
 profiles occurring in the stronger bars and the exponential ones in
 less strong ones.

 Apart from these differences, we find agreement with the results of 
 Miwa \& Noguchi \shortcite{mn98}. Thus, in our purely stellar
 simulations we find a roughly linear correlation between the pattern
 speed of the regenerated bar and its strength, or, accordingly, the
 interaction parameter $\Theta$ (see Fig.~\ref{fig17} and
 ~\ref{fig15}, respectively). In other words, stronger bars --
 regenerated by the interaction -- have lower pattern speeds. This
 result is in agreement with the simulations of Miwa \& Noguchi
 \shortcite{mn98}, who find a similar correlation -- less strict
 though -- between the pattern speed of the induced bars and the
 companions mass. In both cases, i.e. the formation \cite{mn98} and
 regeneration of stellar bars (see Fig.~\ref{fig17}) by tidal
 interactions, the bars always have lower pattern speeds than the
 bars formed spontaneously in the corresponding isolated models.

 In agreement with Noguchi \shortcite{nog87} and Gerin et al.
 \shortcite{ger90} we find that tidal bars are not transient, but
 long-lived, as are spontaneous bars. Thus bars which are
 observed at high redshifts \cite{srss03} may have formed by galaxy
 interactions and thus played an important role in the evolution of
 disc galaxies \cite{srss03}.

\subsection{Numerical considerations}

 The simulations presented in this work have been performed with a
 rather low particle number of the host galaxy, because of the
 relatively large sample of simulations. On the one hand the induced
 numerical noise in general supports the bar formation process, while
 on the other it will heat the disc, and thus make the bar
 regeneration more difficult. We trust, however, that a higher number
 of particles would make only quantitative, and not qualitative
 differences. Thus the physical results should remain unchanged.

 The companion galaxy in our simulations has always been approximated
 by a softened point mass. For simulations in which the companion
 passes outside the halo of the host galaxy, this approximation is
 well suited. For simulations in which the companion crosses the
 halo, however, dynamical friction can be sufficiently strong to
 change the orbit of the companion and finally lead to a merger
 between the two galaxies. 

\section{Summary} \label{summary}

 In this paper we used numerical simulations to investigate the
 regeneration of a stellar bar by tidal encounters. The host galaxy
 has been chosen to be initially bar unstable and forms a
 large-scale bar during its early evolution. Before the interaction
 with the companion galaxy the bar significantly weakened, owing to
 the bar-driven gas-inflow towards the central disc region. For the
 simulations presented in this work we used two different types of
 host galaxies, i.e. one gas-rich disc model and one without gas.
 The encounters in our simulations have been chosen to be prograde
 with respect to the rotation in the host disc and co-planar with the
 discs equatorial plane. The mass and the orbit of the companion
 have been varied in order to cover a large parameter space.

 We found that interactions, which are sufficiently strong to
 regenerate the bar in the purely stellar models, do not lead to a
 regeneration in the dissipative models. The regenerated bars in our
 simulations are long-living phenomena and by no means only
 transient, i.e. that (regenerated) bars formed by interactions
 may indeed contribute to the number of barred spirals at high
 redshifts and in the local Universe. 
 
 We have shown that the strength of the regenerated bars increases
 with the interaction strength. Owing to the tidal perturbation,
 angular momentum is removed from the disc. In fact, the whole
 disc within its initial cut-off radius loses angular momentum,
 in contrast to what is found for isolated discs, where the corotation
 radius of the bar separates disc regions losing and gaining
 angular momentum. The amount of angular momentum removed from the
 disc shows a clear correlation with the interaction strength.
 We argued that the main effect of this angular momentum loss is
 a significant extension of the region where bar-supporting orbits
 exist, resulting in a lengthening of the bar. As a further,
 though somewhat less important, effect, we found also a
 thinning of both the bar-supporting orbits and the bar itself.

 The regenerated bars have lower pattern speeds than the bars in the
 corresponding isolated models. Furthermore, we found strong
 correlation between the strength and the pattern speed of the bar.
 This correlation is in very good agreement with the correlation
 found for bars in isolated discs. This is one of the pieces of
 evidence that the regenerated bars are qualitatively similar to
 those formed in isolated discs and thus cannot easily (if at all)
 be distinguished from them by their dynamical properties.

 In contrast to the purely stellar simulations presented in this
 work, it has not been possible to regenerate the bar in our
 models including gas. We argued that this is due to the fact that
 less angular momentum is lost from the inner disc and to the fact
 that, owing to the interaction, additional gas is driven towards
 the centre of the disc. We concluded that the regeneration of
 stellar bars by galaxy interactions seems to be a reasonable
 mechanism for galaxies containing not too much gas, provided the
 external forcing is sufficiently strong.
 
\section{ACKNOWLEDGMENTS}

 We would like to thank Albert Bosma for interesting discussions, 
 J.C. Lambert and C. Theis for their computer assistance, and the
 referee, Ron Buta, for his comments. 

 I.B. acknowledges support from DFG grant Fr 325/48-2, /48-3 and 
 Volkswagen Foundation grant I/76\,520. He thanks the DFG for funding
 the grape facilities at the Sternwarte G\"ottingen.
 C.H. acknowledges support from grants HST-AR-09546.02-A, NSF-AST-0206251,
 GSU-FRG-2002, and DFG Fr 325/39-1, /39-2.
 E.A. would like to thank the IGRAP, the INSU/CNRS and the University
 of Aix-Marseille {\rm I} for funds to develop the grape facilities
 used for part of the calculations in this paper. The final draft of
 this paper was written while E.A. was in I.N.A.O.E. She thanks the
 I.N.A.O.E. staff for their kind hospitality and ECOS-Nord/ANUIES for
 a travel grant that made this trip possible.
 

\end{document}